\documentclass[useAMS,usenatbib]{mn2e}
\usepackage{graphicx}

\def\arrc{\arraycolsep=2pt}
\def\figref#1{Fig.\,\ref{#1}}
\def\Gyr{\,{\rm Gyr}}
\def\msun{\,{\rm M}_\odot}
\def\dex{\,{\rm dex}}
\def\d{{\rm d}}\def\e{{\rm e}}
\def\kpc{\,{\rm kpc}}
\def\kms{\,{\rm km\,s}^{-1}}
\def\mag{\,{\rm mag}}
\title[History of the solar neighbourhood] %% give here short title %%
{Kinematics and history of the solar neighbourhood revisited}
\author[Michael Aumer \& James J. Binney]{Michael Aumer$^{1}$\thanks{
aumer@usm.lmu.de } and James J.
Binney$^{2}$\thanks{ binney@thphys.ox.ac.uk }\\
$^{1}$Universit\"atssternwarte M\"unchen, Scheinerstr. 1, 81679 M\"unchen, D\\
$^{2}$Rudolf Peierls Centre for Theoretical Physics, University of Oxford, Oxford OX1 3NP, UK}
\begin{document}

\date{Accepted . Received ; in original form}

\pagerange{\pageref{firstpage}--\pageref{lastpage}} \pubyear{2002}

\maketitle

\label{firstpage}

\begin{abstract}
We use proper motions and parallaxes from the new reduction of
\textit{Hipparcos} data and Geneva-Copenhagen radial velocities for a
complete sample of $\sim15\,000$ main-sequence and subgiant stars, and new
Padova isochrones to constrain the kinematics and star-formation history of
the solar neighbourhood. We rederive the solar motion and the structure of
the local velocity ellipsoids. When the principal velocity dispersions are
assumed to increase with time as $t^\beta$, the index $\beta$ is larger for
$\sigma_W$ ($\beta_W\approx0.45$) than for $\sigma_U$ ($\beta_U\approx0.31$).
For the three-dimensional velocity dispersion we obtain $\beta=0.35$.  We
exclude saturation of disc heating after $\sim3\Gyr$ as proposed by
\citet{Quillen}.  Saturation after $\ga4\Gyr$ combined with an abrupt
increase in velocity dispersion for the oldest stars cannot be excluded. For all
our models the star-formation rate is declining, being a factor 2--7 lower now than it was
at the beginning. Models in which the SFR declines exponentially favour very
high disc ages between 11.5 and $13\Gyr$ and exclude ages below
$\sim10.5\Gyr$ as they yield worse fits to the number density and velocity
dispersion of red stars.  Models in which the SFR is the sum of two declining
exponentials representing the thin and thick discs favour ages between 10.5
and $12\Gyr$ with a lower limit of $\sim10.0\Gyr$. Although in our models the
star-formation rate peaked surprisingly early, the mean formation time of
solar-neighbourhood stars is later than in ab-initio models of galaxy
formation, probably on account of weaknesses in such models.
 \end{abstract}


\section{Introduction}\label{sec:intro}

For the solar neighbourhood we have the most detailed observational data
available for any galactic disc. In particular, the distributions within the
solar neighbourhood  of stellar ages,
metallicities and space velocities are keys to
deducing how the disc has evolved  chemically and dynamically. Work directed
at understanding
how galaxies have formed and evolved is a major area of contemporary
astronomy, and studies of the local disc have an important role to play in
this effort. Despite much progress  many questions about the evolution of the
local disc remain open.

The \textit{Hipparcos} catalogue provides uniquely useful data for studies of
the solar neighbourhood, because the data are homogeneous and of high
quality.  \citet[][hereafter DB98]{DB98} defined a kinematically unbiased
sample of \textit{Hipparcos} stars and investigated its kinematics, while
\citet[][hereafter BDB00]{BDB00} used the sample to model the history of star formation and the
stochastic acceleration of stars in the disc. In this paper we redefine the
sample and then re-work these papers for several reasons.

\begin{itemize} 
\item[(i)] In a systematic re-reduction of the astrometric data from the
\textit{Hipparcos} satellite, \cite{van Leeuwen} has been able to diminish
significantly the errors for a large number of
stars. Using the new data we can enlarge the sample of stars
that can be used for modelling.

\item[(ii)] BDB00 did not take into account the variation of the scale height of
stars with age. Consequently, their conclusions regarding past star-formation
rates and rates of stochastic acceleration are wrong. Our analysis remedies
this defect.

\item[(iii)] From an analysis of the age-velocity dispersion relation in the
sample of 189 stars in \cite{Edvardsson93}, \citet{Quillen} argued that
stochastic heating saturates after only $3\Gyr$ and that the velocity
dispersion of disc stars increased abruptly  $\sim9\Gyr$ ago as a result of a
merger. We use our much larger sample to test this conjecture.

\item[(iv)] Metallicity and radial-velocity measurements from the Geneva
Copenhagen Survey [\citet{GCS}, hereafter GCS, \citet{GCS2}, hereafter
GCS2 and \citet{GCS3}, hereafter GCS3] of F and G dwarfs allow more accurate determinations of the metallicity
distribution of disc stars and  the
three-dimensional velocity dispersions in the corresponding colour interval.

\item[(v)] New Padova isochrones have recently been published  \citep{Bertelli},
allowing us to update the input concerning stellar evolution and to improve
the modelling of the metallicity distribution.

\item[(vi)] A revision of the impact of interstellar reddening on 
our sample  changes some
results significantly.

\end{itemize}

The goal of this paper is the same as that of BDB00: to understand as much of
the history of the disc as we can by modelling the large-scale structure of
velocity space near the Sun. We do not address the small-scale structure of
this space, which was first clearly revealed by the \textit{Hipparcos} Catalogue
through the work of \cite{Creze98} and \cite{Dehnen98}.

Section \ref{sec:data} describes our input data. Section
\ref{sec:kinematics} updates DB98 by
extracting the kinematics of the solar neighbourhood.  The work of BDB00 is
updated in Section \ref{sec:models}, which
describes our models, and Section \ref{sec:fits}, which describes fits of them to
the data.  Section \ref{sec:conclude} sums up and compares our results
with those of other authors.

\begin{figure}
\hspace{-1.cm}
 \scalebox{0.7}{\includegraphics[angle=270]{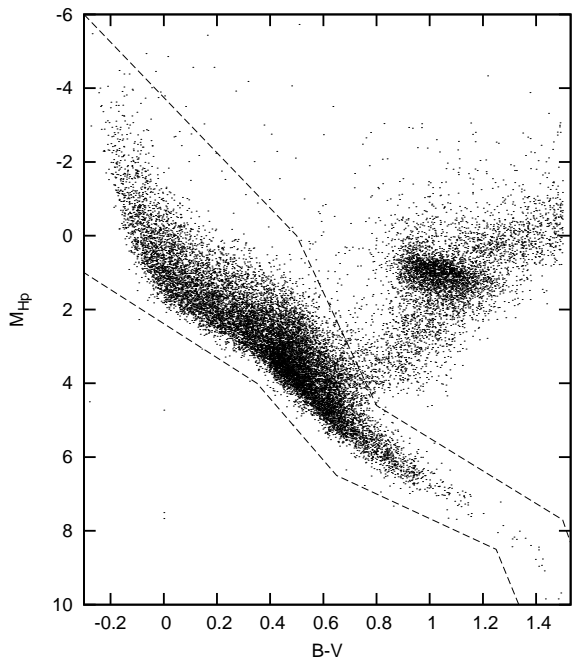} }
  \vspace{ 1.4cm}
 \caption{Colour-magnitude diagram of the magnitude limited subsample of
 20\,360 \textit{Hipparcos} stars with relative parallax errors smaller than
 10\%. The lines indicate our selection of main-sequence stars, there are 15\,113 stars between the lines}
   \label{fig1}

\end{figure}

\section{The Data}\label{sec:data}
\subsection{The Sample}\label{subsec:sample}

We follow the procedure of DB98 to select from the new reduction of the
\textit{Hipparcos} catalogue \citep{van Leeuwen} a kinematically unbiased,
magnitude-limited subsample of single stars with high-quality astrometric
data.

We determine the apparent magnitude up to which the \textit{Hipparcos}
catalogue is complete ($V_{\rm{lim}}\approx{8}\,$mag), which depends on
colour and position on the sky, by comparing the \textit{Hipparcos} catalogue
with the
\textit{Tycho2} catalogue \citep{Tycho2}, which is complete to about
$V_{\rm{Tycho}}=11$.  For $16\times16\times10$ uniformly spaced bins in
galactic coordinates $\sin b$ and $l$ and colour
$(B-V)_{\rm{Tycho}}$ in $(-0.3,1.5)$, we select those stars
that are brighter than the second brightest star per bin that is in the
\textit{Tycho2} but not the \textit{Hipparcos} catalogue. 44$\,$567 out of the
118$\,$218 \textit{Hipparcos} stars are single stars that fulfil this
criterion.  From these stars we further select stars with relative parallax
errors of $10\%$ or less. This criterion reduces the sample from 44$\,$567 to
20$\,$360 but ensures that the measured proper motions yield fairly accurate
tangential velocities.

 Fig.\,\ref{fig1} is the colour-magnitude diagram of the sample.  For our
analysis of the star-formation history, we require a one-to-one relationship
between the maximum lifetime of a star and its colour, so we restrict
ourselves to main-sequence stars. The lines within which stars are deemed to
lie on the main sequence are shown in Fig.\,\ref{fig1}. The applied CMD cut leaves the possibility
for a small number of subgiant stars to enter the sample in the colour range
$0.6<B-V<0.75$, where the sample might thus be slightly biased towards old stars.
The 15$\,$113 stars that lie between these lines comprise our final sample, which is
approximately 27\% larger than that of DB98. The growth in the sample is
most pronounced for blue stars since these tend to be more luminous and
distant and therefore have the smallest parallaxes.

GCS have measured radial velocities for 6$\,$918 single F and G dwarfs in our
sample. These stars are confined to the colour interval $0.4<B-V<0.8$.  We
use this subsample, for which individual space velocities can be determined,
as a control of the results obtained with the main sample.

\subsection{Isochrones and Metallicities}\label{subsec:isochrones}

The age distribution of main-sequence stars of a given colour must vary with
colour because at the blue end of the main sequence all stars must be younger
than the short main-sequence lifetime there, while at the red end of the main
sequence we see stars that are as old as the disc in addition to
recently-formed stars.  We use Padova isochrones \citep{Bertelli} for masses
ranging from 0.15 to $4\msun$ to determine the age distribution at each
colour.
Isochrones depend significantly on the metal content $Z$ and the Helium
abundance $Y$, so we have to use several isochrones to simulate the
metallicity distribution of the solar neighbourhood. We refer to Section
\ref{subsec:Z} for a discussion of the weightings of the isochrones employed and
focus on which isochrones to employ.

For the conversion from solar units to $Z$ and $Y$ we need to know the solar
abundances. \citet{Bertelli} argue that the update of the solar chemical
composition by \citet{Grevesse07} ($Y_{\odot}=0.2486$,$Z_{\odot}=0.0122$)
gives rise to several problems and uncertainties and adopt
($Y_{\odot}=0.260$,$Z_{\odot}=0.017$) for their solar model. Moreover
\citet{Chaplin07} report that solar models with $Z<0.0187$ are inconsistent
with the helioseismic data. Since we use the \citet{Bertelli} isochrones, we
use their solar composition.

As the measurement of the Helium abundance $Y$ of a star is difficult, there
is little data available on the relationship $Z(Y)$ within the solar
neighbourhood. A common, but not undisputed procedure is to assume a linear
enrichment law $Y=Y_{\rm p}+(\d{Y}/\d{Z})Z$, where $Y_{\rm p}$ is the
primordial helium abundance and $\d Y/\d Z$ is the helium-to-metal enrichment
ratio. \citet{Jimenez} find $\d{Y}/\d{Z}=2.1\pm{0.4}$ for $Y_{\rm p}=0.236$,
whereas \citet{Casagrande} using Padova isochrones find that it is doubtful
whether a universal linear enrichment law exists. For metallicities similar
to the solar one they also find $\d{Y}/\d{Z}=2.1\pm{0.9}$.  Unfortunately,
this slope when combined with either the traditional values
$(Y_\odot,Z_\odot)=(0.260,0.017)$ or the new values $(0.2486,0.0122)$ of
\citet{Grevesse07} is inconsistent with the generally accepted WMAP value $Y_{\rm
p}=0.24815$ \citep{Spergel}. Nonetheless we adopt $\d{Y}/\d{Z}=2.1$.
Table\,\ref{tab8} gives the chemical compositions of the isochrones we have
employed.

\begin{table}
	\centering
	\caption{The metallicities of the Padova isochrones that were used in
the models and the corresponding $\rm[Fe/H]$ values, which were calculated
with solar abundances $Z_{\odot}=0.017$ and $Y_{\odot}=0.260$.}
  \label{tab8}
		\begin{tabular}{cc|r}
\hline
 $Z$&$Y$&[Fe/H]\\

\hline
0.002 & 0.23 & $-0.95$\\
0.003 & 0.23 & $-0.78$\\
0.004 & 0.23 & $-0.65$\\
0.006 & 0.24 & $-0.47$\\
0.008 & 0.24 & $-0.34$\\
0.010 & 0.25 & $-0.24$\\
0.012 & 0.25 & $-0.16$\\
0.014 & 0.25 & $-0.09$\\
0.017 & 0.26 & 0.00\\
0.020 & 0.27 & 0.08\\
0.026 & 0.28 & 0.20\\
0.036 & 0.30 & 0.36\\
\hline
		\end{tabular}
\end{table}

\begin{figure*}
\hspace{-8.5cm}
\scalebox{0.8}{\includegraphics[angle=270]{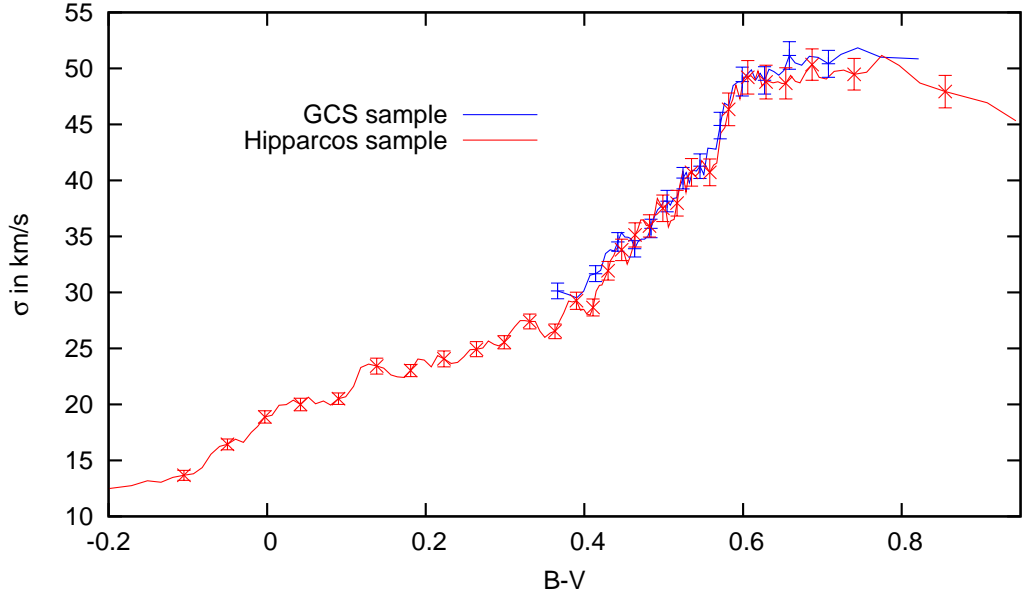} }
  \vspace*{-5.7 cm}
 \caption{Total velocity dispersion vs.\ colour. The red line connects the data points for the
van Leeuwen--\textit{Hipparcos} proper motions and here and elsewhere only
independent data points carry 1$\sigma$-errorbars. The blue line is the
relation obtained from the GCS space velocities. }
   \label{fig2}

\end{figure*}

\section{Stellar kinematics}\label{sec:kinematics}

We follow the method of DB98 to calculate the three-dimensional velocity
dispersion $\sigma$ as a function of colour and use a sliding window in $B-V$
with borders that were adjusted to ensure that there are always 500 stars in
a window. Every 100 stars a new point is plotted and thus every fifth point
is statistically independent of its predecessors. As explained in DB98, the
averages we use here are sensitive to outliers, which have to be rejected. We
therefore use an iterative method and reject stars that contribute to
$\sigma^2=\sigma_U^2+\sigma_V^2+\sigma_W^2$ more than $\kappa^2$
times the value of the previous iteration.\footnote{We use a right-handed
coordinate system with $x$ increasing towards the Galactic centre and $y$
increasing in the direction of Galactic rotation.} We use $\kappa=3.5$ and 6
iterations; 55 stars were rejected.

The red points in Fig.\,\ref{fig2} show the resulting run of $\sigma$ against
the mean $B-V$ colour in the sliding windows, with errors in $\sigma$ shown
for every fifth point.  Parenago's discontinuity -- the abrupt end to the
increase in velocity dispersion with $B-V$ \citep{Parenago} -- is beautifully
visible at $B-V\approx{0.61}$. We also calculated $\sigma$ by combining GCS
radial velocities with Hipparcos proper motions, and in \figref{fig2} the
results are plotted in blue. Where the samples are comparable, there is
excellent agreement between the two measurements of $\sigma$.  At the upper
and lower limits of the GCS sample there are discrepancies because the GCS
stars were selected by spectral type, so the bluest GCS bins only contain F
stars, while the bins from the \textit{Hipparcos} sample contain A and F
stars and consequently have a lower mean age. For the reddest bins the
situation is the same for G and K stars, but this time counterintuitively the
pure G star sample has a higher mean age and velocity dispersion -- we
explain this phenomenon in Section \ref{subsec:reddrop}.

\begin{figure}
\hspace{-2cm}
\scalebox{0.9}{\includegraphics[angle=270]{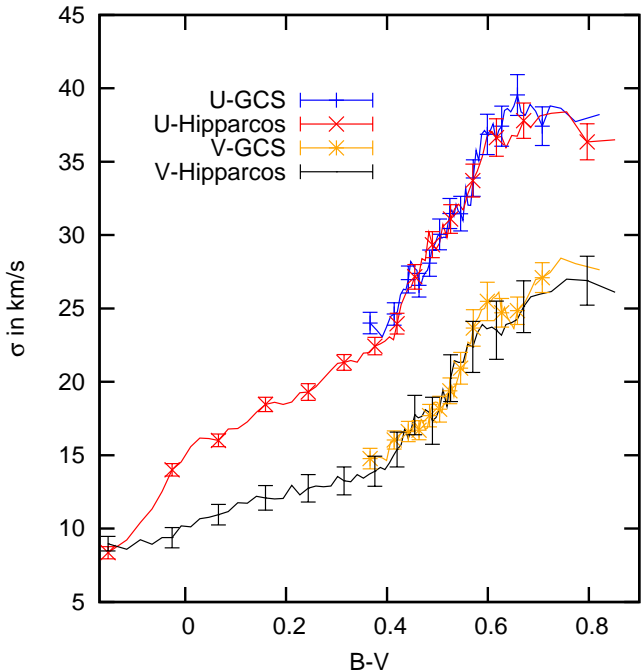} }
  \vspace*{0.6 cm}
 \caption{In-plane components of velocity dispersion vs.\ colour. The red line connects the
data points for the $\sigma_U$ component and the black line connects the data points for the
$\sigma_V$ component, both calculated from the van
Leeuwen--\textit{Hipparcos} proper motions with bins of 1000 stars. We also
show $\sigma_U$ and $\sigma_V$ from the GCS space velocities with bins
of 500 stars. Only independent data points carry 1$\sigma$-errorbars.}
   \label{fig3}

\end{figure}

\begin{figure}
\hspace{-2cm}
\scalebox{0.9}{\includegraphics[angle=270]{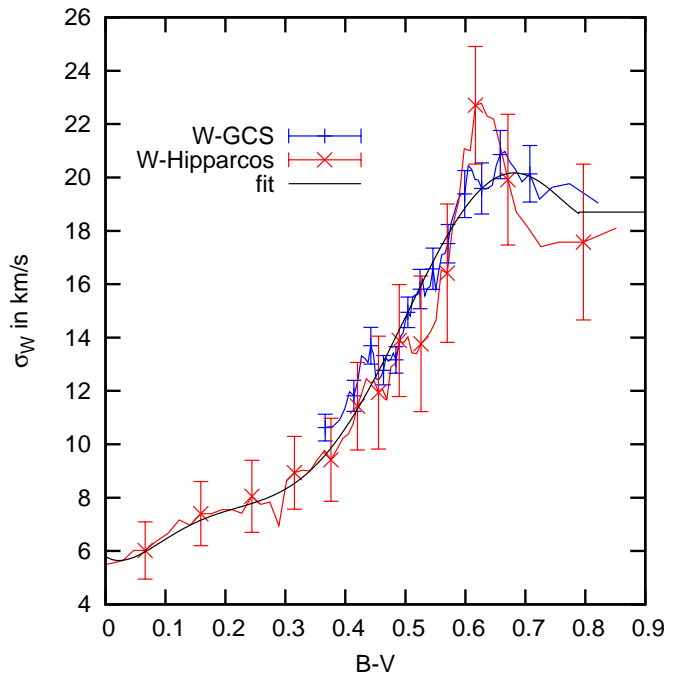} }
  \vspace*{0.3 cm}
 \caption{Vertical velocity dispersion vs.\ colour. The red line connects data points
for $\sigma_W$ from the van Leeuwen--\textit{Hipparcos} proper
motions with bins of 1000 stars, while the blue points show values from the GCS
space velocities with bins of 500 stars. Only independent data points carry 1$\sigma$-errorbars.
The black line shows the polynomial
fit that is explained in the text and used in subsequent work.}
   \label{fig4}

\end{figure}

The diagonal elements of the velocity-dispersion tensor are also of interest,
as the stochastic acceleration mechanisms at work are anisotropic so each
component of the tensor can evolve independently in time (cf. e.g. GCS).
Fig.\,\ref{fig3} plots $\sigma_U$ and $\sigma_V$ against the mean $B-V$
colour of each bin and Fig.\,\ref{fig4} shows $\sigma_W$. The
three-dimensional dispersion is dominated by the radial dispersion
$\sigma_U$, which consequently shows the same features as $\sigma$. The
other two components are different, however, and Parenago's discontinuity is
not as beautifully visible in them. For $\sigma_V$, the dispersion in the
direction of rotation, we notice that the slope changes at $B-V\approx{0.6}$,
but it continues to increase until $B-V\approx{0.75}$. Interestingly, in
\figref{fig4} the values $\sigma_W$ from proper motions (red points) show
a significant bump at $B-V\approx{0.6}$. A similar bump is visible in DB98
but less strikingly so on account of the larger bins used by DB98.  Values of
$\sigma_W$ from GCS space velocities have smaller errors and show a less
distinctive bump. The  GCS error bars on individual components of velocity
dispersion are significantly smaller than the corresponding error
bars from the larger Hipparcos sample
because each Hipparcos star in effect constrains only two components of
velocity dispersion. The error bars from the two measurements of $\sigma_W$
overlap, so the results are not inconsistent. The bump originates from a 
decrease in average age redwards of the discontinuity that we will explain
in Section \ref{subsec:reddrop}. In subsequent work we require a
functional fit to the data in \figref{fig4}. Since the radial velocities add
valuable additional information, we fit a 5th-order polynomial to the GCS
values where these are available, and elsewhere to the proper-motion values.
The black curve in Fig.\,\ref{fig4} shows the resulting relation.

\begin{figure} 
\hspace{-2.2cm}
\scalebox{0.9}{\includegraphics[angle=270]{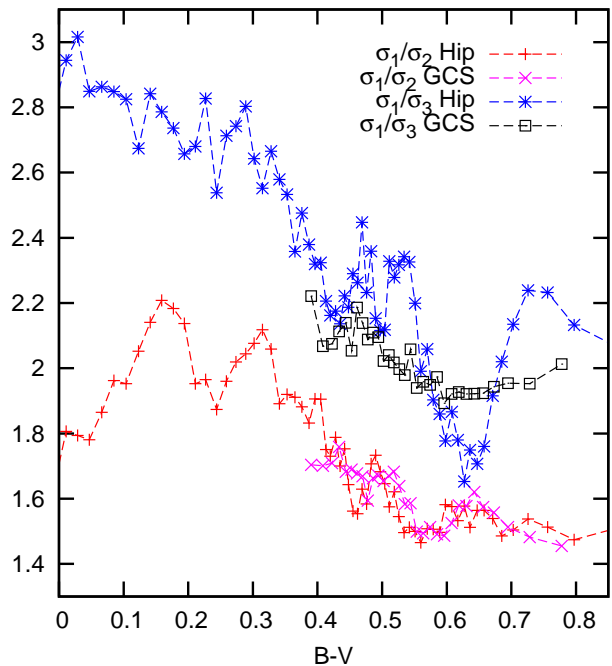} }
  \vspace*{0.3 cm}
 \caption{The ratios of the eigenvalues of the velocity dispersion tensor for
bins with 1000 stars and a new bin after 200 stars.
${\sigma_1}/{\sigma_2}$ from proper motions (red) and GCS space
velocities (magenta). ${\sigma_1}/{\sigma_3}$ from proper motions (blue)
and GCS space velocities (black).}
   \label{fig10}
\end{figure}

As in DB98, the $\sigma_{UV}^2$ component of the velocity dispersion tensor
is nonzero, implying that the principal axes of $\sigma^2$ are not aligned
with our coordinate axes. We diagonalised $\sigma^2$ to find its eigenvalues
$\sigma_i^2$, the ratios of the square roots of which are plotted in
Fig.\,\ref{fig10}. The ratio ${\sigma_1}/{\sigma_2}$ (red and magenta) shows
a slight decrease with increasing $B-V$ and therefore age, dropping from
$\sim1.9$ to $\sim1.6$ through the range $0.3<B-V<0.55$.  The ratio
${\sigma_1}/{\sigma_3}$ drops from $\sim2.8$ for the bluest stars to $\sim2$
(judging from the GCS sample) for stars redder than Parenago's discontinuity.

We calculated the sun's velocity with respect to the LSR $(U_0,V_0,W_0)$ as
explained in DB98 and found:
{\arrc
\begin{eqnarray}
U_0&=&(9.96\pm0.33)\,\rm{km\,s^{-1}}\nonumber\\
V_0&=&(5.25\pm0.54)\,\rm{km\,s^{-1}}\\
W_0&=&(7.07\pm0.34)\,\rm{km\,s^{-1}}\nonumber.
\end{eqnarray}
}These values are  consistent with the results of DB98 but our error
bars are smaller by $\sim11$ per cent. \citet{van Leeuwen}, using a sample of
$\sim20\,000$ main-sequence stars selected by imposing only an
upper limit for relative parallax errors of 10\%, found $U_0=10.77\pm0.36$,
$V_0=3.21\pm0.52$, in conflict with our values, and $W_0=7.04\pm0.14$, which is 
consistent with our findings. Since van Leeuwen's sample extends below the
limit to which the \textit{Hipparcos} sample is photometrically complete,
his solar motion may be affected by the kinematic biases that are known to be
present in the \textit{Hipparcos} input catalogue: stars thought be
interesting for a variety of reasons were added to the catalogue, and many such
stars had come to the attention of astronomers by virtue of a high proper
motion.

Using our values for the solar motion and for $\sigma^2_{xx}$
to calculate the coefficient in Str\"omberg's
asymmetric drift equation
\begin{equation}
\langle v_y \rangle = - \sigma^2_{xx}/k,
\end{equation}
we find $k=74\pm5\kms$, which is lower than, but still consistent with the
result of DB98, and also with  the estimated scale length of the Galactic
disc \citep[][\S4.8.2(a)]{GD2}.

\begin{figure} 
\hspace{-0.4cm}
\scalebox{0.9}{\includegraphics[angle=270]{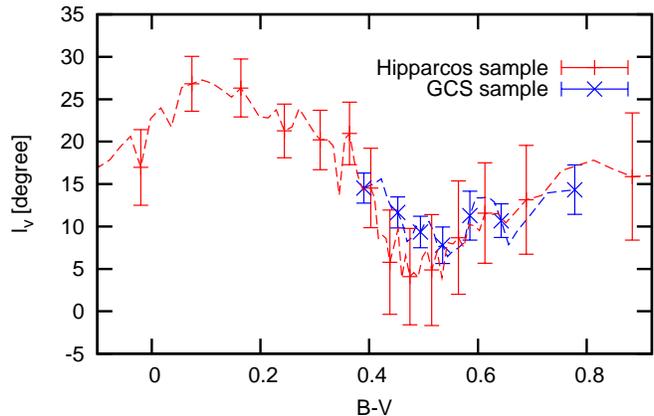} }
  \vspace*{0.3 cm}
 \caption{The vertex deviation $l_v$ as a function of colour $B-V$ for bins containing 1000
 stars each.}
   \label{fig30}
\end{figure}

As in DB98, we also calculated the vertex deviation $l_v$, the angle by which
one has to rotate the applied Cartesian coordinate system around its
$\hat{\textbf{z}}$ axis to make the velocity dispersion tensor diagonal in
the $(v_x,v_y)$ plane. Fig.\,\ref{fig30} shows the vertex deviation as a
function of colour $B-V$ for bins containing 1000 stars each.  As in DB98 the
vertex deviation has a minimum at the colour of Parenago's discontinuity. The
displayed errors for the Hipparcos sample are slightly larger than in DB98
because we use smaller bins.

\section{Modelling the history of star formation and
heating}\label{sec:models}

\subsection{Interstellar reddening}\label{subsec:reddening}

For our models, it is important to apply a correction for interstellar reddening 
to the data because reddening shifts the
location of Parenago's discontinuity, which strongly influences the
results of our models for the star formation history. BDB00 corrected the
data assuming a linear reddening law, $E(B-V)=0.53\mag\kpc^{-1}$. This law is
appropriate for distances of order kiloparsecs, but we live in a low-density
bubble within the ISM, so the column density of the ISM
and thus the amount of reddening depends on the direction of the line of
sight \citep[e.g.][]{Frisch} and for small distances is typically smaller
than the mean relation would imply. The code of \citet{Hakkila} allows us to
determine the reddening as a function of galactic coordinates $(l,b)$, but
unfortunately cannot reproduce the results of \citet{Lallement} or
\citet{Vergely}, who find that the reddening within $\sim70\,$pc of the sun
is essentially negligible. Recently GCS2 found that reddening is negligible
within $40\,$pc and $E(b-y)=0.0048\mag$ between 40 and $70\,$pc.\footnote{The
conversion factor between colour systems is $E(B-V)\approx1.35\,E(b-y)$
\citep{Persinger}, so GCS2 find $E(B-V)\simeq0.0065$ between 40 and $70\,$pc.}

   \begin{figure}
\hspace{-1.5cm}
\scalebox{0.9}{\includegraphics[angle=270]{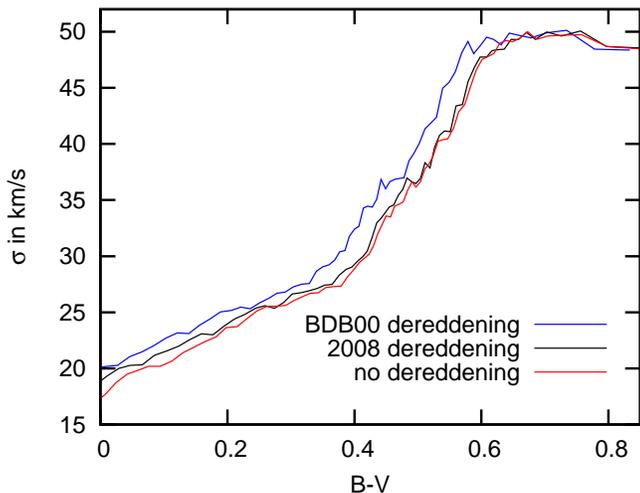} }
  \vspace*{-2 cm}
 \caption{The effect of dereddening on the $\sigma$ vs. $(B-V)$ diagram for
bin sizes of 1000 stars.}
   \label{fig5}
   \end{figure}
   
We decided to deredden according to Fig.~4 of \citet{Vergely}. For stars
within $40\,$pc of the midplane and with a distance from the sun between
$d=70$
and $300\,$pc, they find $E(b-y)\approx0.35(d-70\,\rm{pc})/\rm{kpc}$, so we
have used
\begin{equation}
E(B-V)=\cases{0&for $d<70\,\rm{pc}$\cr
0.47(d-70\,\rm{pc})/1\kpc&otherwise.
}
\end{equation}

Fig.\,\ref{fig5} displays the effect of dereddening on the $(B-V,\sigma)$
diagram by comparing data dereddened according to the old and new
prescriptions. The binsize used here was 1000 stars and every 200 stars a new
data point was added. We see that BDB00 considerably overestimated the effect
of reddening.

\subsection{Volume completeness}\label{subsec:completeness}

As the velocity dispersion vs. colour $C\equiv B-V$ diagram alone allows models with a
rather large variety of parameters, we use  the number of
stars per colour interval, ${\d N/\d C}$, as an additional constraint. Our
model of the dynamics will predict $N_{\rm cyl}(C)$, the number of stars of a
given colour in a vertical cylinder through the disc that has radius
$R_{\rm{cyl}}$ and the sun on its axis. Two points have to be borne in mind
when relating this to the observed number density $\d N/\d C$: a) the scale
height of stars varies with velocity dispersion and thus colour, and b) the
radius of the sphere within which the sample is complete varies with colour.

Assuming the stellar distribution is in equilibrium and neglecting variations
with galactic radius, the stellar number density as a function of vertical coordinate
$z$, potential $\Phi(z)$ and vertical velocity dispersion $\sigma_W(C)$ is
 \begin{equation}\label{nuofz}
\nu(z, C)=\nu_0(C)\, \exp\left(-\frac{\Phi(z)}{\sigma_W^2(C)}\right),
\end{equation}
 where the normalising factor  $\nu_0(C)$ is to be determined from the
observed density $\d N/\d C$.
Integrating $\nu(z,C)$ through the cylinder we have
 \begin{equation}\label{defsNcyl}
N_{\rm{cyl}}(C)=\nu_0(C) \pi R_{\rm{cyl}}^2 \int_{-\infty}^{\infty} \d z\,
\exp\left(-\frac{\Phi(z)}{\sigma_W^2(C)}\right).
\end{equation}
 Integrating $\nu(z,C)$ through the sphere within which the star count is
 complete at the given colour we have
{\arrc \begin{eqnarray}\label{dNdCofnu}
{\d N\over\d C}&=&\nu_0(C) \pi \int_{-R_c}^{R_c}\!\!\d z\, 
(R_c^2(C)-z^2)\exp\left(-\frac{\Phi(z+z_0)}{\sigma_W^2(C)}\right)\\
&\equiv& \nu_0(C) I(C),\nonumber
\end{eqnarray}
}where $R_c(C)$ is the completeness radius and $z_0$ is the vertical distance of the sun from the galactic midplane.
We now have that the normalising constant $\nu_0$ is
 \begin{equation}\label{givesnuzero}
\nu_0(C)=\frac{\d N/\d C}{I(C)}
\end{equation}
 and on substituting this into equation (\ref{defsNcyl}) we obtain the
required relation between $N_{\rm cyl}$ and the star counts $\d N/\d C$.
$R_{\rm{cyl}}$ can be set to any value larger than the largest value of
$R_c(C)$. We set $z_0=15\,$pc \citep{BinneyGS,Joshi} and for $\sigma_W(C)$ we
use the polynomial fit shown in \figref{fig4}. For the potential
$\Phi(z)$ we use Model 1 of \S2.7 of \citet{GD2}. 

\begin{figure}
\hspace{-1.7cm}
\scalebox{0.9}{\includegraphics[angle=270]{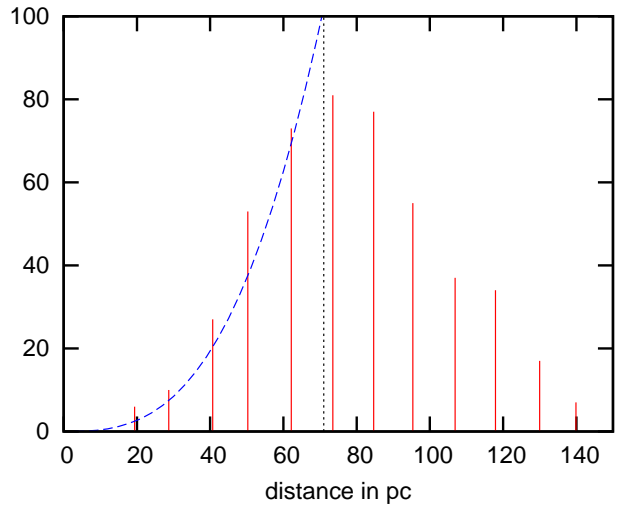} }
  \vspace*{-1.8 cm}
 \caption{Distance histogram for the colour bin with $0.459<B-V<0.476$. The dashed line shows the expected relation for a volume complete sample. The final radius used was 71pc and is indicated by the dotted line.}
   \label{fig7}
   \end{figure}

\begin{figure}
\vspace*{-0.5 cm}
\hspace{-1.5 cm}
\scalebox{0.9}{\includegraphics[angle=270]{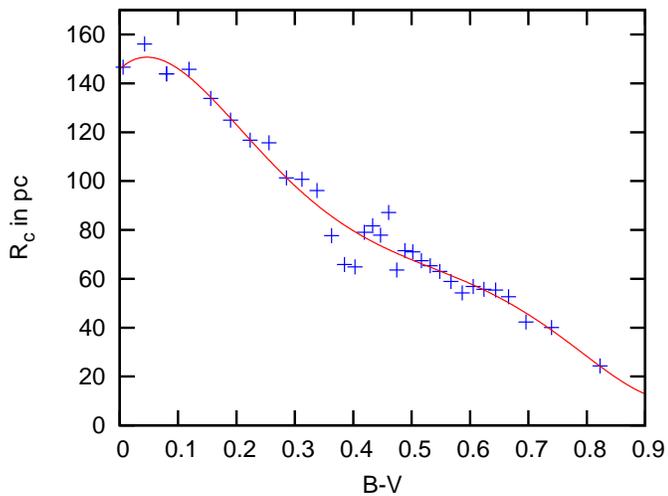} }
  \vspace{-2.1 cm}
 \caption{The blue data points show $R_c$ as determined by the Kolmogorov-Smirnov test. The red line displays the applied 5th order polynomial fit.
}
   \label{fig8}
   \end{figure}

For the determination of $R_c$ we compare the distribution of distances of
stars within a radius $R$ in a given colour bin to the distribution we would
expect, namely
 \begin{equation}\label{defsncyl}
n(R,C)=\nu_0(C) \pi \int_{-R}^{R}\!\!\d z\, 
(R^2-z^2)\exp\left(-\frac{\Phi(z+z_0)}{\sigma_W^2(C)}\right)
\end{equation}
 The model density $n(R,C)$ depends on $R_c$ through $\nu_0(C)$ and we adjust
$R_c$ until we have a value $0.4<P<0.6$ for the Kolmogorov--Smirnov (KS)
probability that the observed distance distribution at $R<R_c$ is consistent
with the model distribution.  These comparisons were made for colour bins
that contain 500 stars each. Fig.\,\ref{fig7} shows a typical histogram and
the corresponding model distribution (\ref{defsncyl}). We finally applied a
5th order polynomial fit to the relation $R_c(C)$ (cf. Fig.\,\ref{fig8}) and
used these radii in eqs.~(\ref{defsNcyl}) and (\ref{givesnuzero}) to relate
$N_{\rm cyl}$ to $\d N/\d C$. An average of 49 per cent of the stars in a
colour bin lie within $R_c(C)$.

A priori, it is not clear, how to choose the exact limits for the KS
probability. \figref{fig7} might give rise to doubts concerning the
completeness out to the chosen value of $R_c$. However, if we require higher
KS probabilities resulting in lower completeness radii and smaller numbers of
stars available for analysis, the results do not change in any
significant way.  We estimated the errors in the number of stars in the
column by varying $R_c$ by 10 per cent around our preferred value and by
varying $\sigma_W$ by its $1\sigma$-errors. The resulting error was added in
quadrature with the Poisson error.  This procedure yielded relative errors
that varied from below 10 to 25 per cent, but we imposed a lower limit of 15
per cent on the relative error.

\begin{figure}
\hspace{-1.55cm}
\scalebox{0.9}{\includegraphics[angle=270]{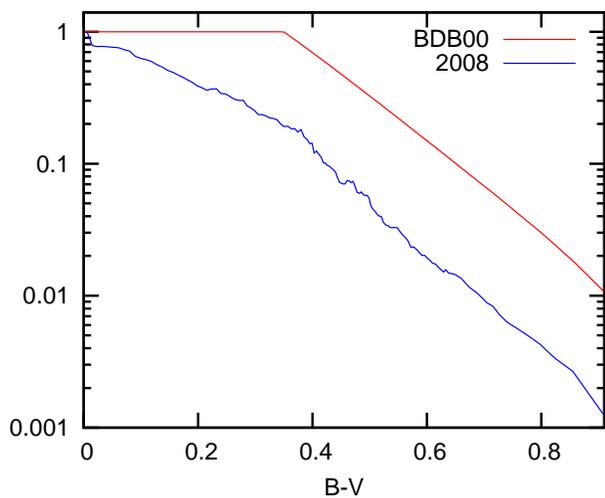} }
  \vspace*{-2. cm}
 \caption{The selection function as in BDB00 (red) and in our models (blue)
}
   \label{fig9}
   \end{figure}

To relate all this to BDB00, Fig.\,\ref{fig9} shows our selection function
(blue) and that of BDB00 (red). One fundamental difference between BDB00 and
this work is that we consider a column stretching to $\infty$, whereas BDB00
considered a sphere with a radius of $100\,$pc. In BDB00 stars within this volume 
more luminous than $5L_\odot$  were assumed to be complete, and for less
luminous stars the radius of completeness was supposed to decrease as
$L^{{1}/{2}}$. Hence in \figref{fig9} the BDB00 selection function is
unity bluer than $B-V\sim0.35$ and then declines linearly.  Since we consider
a column that extends to infinity, the selection function is always a
declining function of luminosity and therefore colour.  Hence the most
significant difference between the two selection functions is at the blue end
where (a) the ability of luminous stars to enter the sample even when far
from the plane makes them relatively more numerous in the sample, and (b) the
more accurate reduction of the raw \textit{Hipparcos} data by \citet{van
Leeuwen} has made stars with higher distances available. Even at fainter
magnitudes the introduction of the scale-height correction has slightly
increased the slope of the selection function, thus depressing the chances of a
red star to enter the sample. Consequently, our models have to increase the
predicted numbers of red stars in a cylinder relative to the predictions of
the BDB00 models.

In summary, roughly half the sample stars contribute to the values of $N_{\rm
cyl}$ that we model, but all of the sample stars contribute to the modelled
values of the velocity dispersions.

\subsection{Distribution over age of stars at a given
colour}\label{subsec:agedistrib}

Following BDB00 the distribution of main-sequence
stars in a certain volume over age and mass is given by:
\begin{equation}
 \frac{\d^2 N}{\d M\d\tau}\propto   \cases{
                   \xi(M)\, \hbox{SFR}(\tau) & for
		   $\tau<\tau_{\rm{max}}(M)$,\cr
                   0 & otherwise,}
\end{equation}
where $\tau_{\rm{max}}(M)$ denotes the main-sequence lifetime of a star of initial
mass $M$, $\xi(M)$ the initial mass function and $\hbox{SFR}(\tau$) the star formation rate.

BDB00 used a Salpeter-like power-law IMF $\xi(M)\propto M^{\alpha}$ and an 
exponential $\hbox{SFR}\propto \exp(\gamma\tau)$. They found
that the corresponding characterictic parameters $\alpha$ and $\gamma$ were strongly
correlated \citep[see also][]{Haywood}. A high, positive value of
$\gamma$, i.e.\ a higher SFR in the past, relatively increases the number of
red stars compared to blue stars. This effect can be cancelled by a relatively
flat IMF creating more blue stars.

Because of this correlation, we decided to use the IMF of \citet{KTG}:
\begin{equation}
 \xi(M) \propto   \cases{
                   1.84\,M^{-1.3} & if $0.08\msun<M<0.50\msun$ \cr
		               \hspace{0.6cm} M^{-2.2} & if $0.50\msun<M<1.00\msun$ \cr
		               \hspace{0.6cm} M^\alpha & if $1.00\msun<M<\infty$.}
\end{equation}
The power-law parameter $\alpha$ for $M>1.0\msun$ has the strongest influence
on our results and is thus allowed to vary around $\alpha=-2.7$, the value
found by \citet{KTG}.

With the IMF fixed within a small range, we tested several models for the star formation 
history:
\begin{itemize} 
\item[(i)] A simple exponential SFR
\begin{equation}
\hbox{SFR}(\tau) \propto \exp(\gamma\tau)
\label{SFRexp}
\end{equation}
\item[(ii)] A SFR of the form
\begin{equation}
\hbox{SFR}(\tau) \propto A\,\exp(\lambda\tau)+\exp(\gamma\tau)
\label{SFRexp2},
\end{equation}
with $\lambda>\gamma$, adding an additional amount of star formation in the early universe.
\item[(iii)] A SFR of the form
\begin{equation}
\hbox{SFR}(\tau)\propto\frac{\tau_2-\tau}{[(\tau_2-\tau)^2+b^2]^2}
\label{SFRJJ},
\end{equation}
as proposed by \citet{JJ}.
\item[(iv)] A smooth SFR overlaid with a factor varying with time
according to Fig.~8 of \citet{RP}.
\end{itemize}

We need the isochrones that were described in Section
\ref{subsec:isochrones} to
determine the mass range that can be found in a colour interval at a given
time. We cut each isochrone off above the point where it is $1.8\mag$ more
luminous than the ZAMS at the same colour. The isochrones provide information
only at a limited number of times, so at a desired time one has to
interpolate between the next older and next younger isochrones; for the
details of this we refer to BDB00. 

We can relate $N_{\rm cyl}$ to the average distribution over age in a colour interval
$(C_-,C_+)$ by
 \begin{equation}\label{xx}
\left \langle{\d N\over\d\tau}\right \rangle_{(C_-,C_+)}= \frac{1}{ C_+-C_-} 
\frac{\exp(\gamma\tau)}{(1+\alpha)} \sum_{j}
(M_{+,j}^{1+\alpha}-M_{-,j}^{1+\alpha}).
\end{equation}
 Here the sum is over all mass ranges $(M_{-,j},M_{+,j})$ that lie in the
colour interval at age $\tau$.

\subsection{Age--metallicity relation}\label{subsec:agemet}

As we use isochrones with a significantly higher number of metallicities
than in BDB00, we are able to include a variation of metallicity with age.
As guidelines we use the metallicity distribution published in GCS2 and the
age--metallicity relation from the models of \citet[][hereafter SB09a]{Ralpha}, who are able
to reproduce the findings of GCS2 and whose results are also consistent
with the age--metallicity relation of \citet{Haywood08}.

We find that the GCS2 metallicity distribution has mean $\overline{\hbox{[Fe/H]}}=-0.12$
and the dispersion $0.17\dex$.  This result is similar to that of
\citet{Girardi}, who found that the metallicity distribution for K giants is
well represented by a Gaussian with a mean of $-0.12$ and a dispersion of
$0.18\dex$. \citet{H01} proposed that the distribution was centred on solar
metallicity, which shows that the uncertainty is not negligible. 
Using a Gaussian representation however omits the metal-poor tail of the
GCS2 distribution, which comprises approximately 4\% of the total sample and
spreads in $\hbox{[Fe/H]}$ from $-1.2$ to $-0.4$.

From the models of SB09a we extracted that stars with $\hbox{[Fe/H]}<-0.7$
form only in the first $\sim 0.6\Gyr$, stars with $-0.7<\hbox{[Fe/H]}<-0.4$ show a 
high contribution from the time interval $0.5-1.5\Gyr$, but also a younger component
and stars with $\hbox{[Fe/H]}>-0.4$ started forming after $\sim0.6\Gyr$ and still form 
today.

We therefore construct a model that features the following two components:

\begin{itemize}
\item[(I)] The `thin disc' component
\vspace{0.1cm}

This component comprises $\sim96\%$ of the model stars and is represented by a 
Gaussian distribution with the above mentioned mean, so its stars belong to  
the isochrones with the ten highest metallicities from 
Table\,\ref{tab8}.
The intrinsic distribution of metallicities will be narrower than the observed one on account of
observational errors (cf. Section \ref{subsec:Z}). GCS2 conclude that there
is no significant change with age in the mean metallicity
of solar-neighbourhood stars, so in our models we can consistently use a fit
to the current distribution at all times. The models for the star-formation history
as described in Section \ref{subsec:agedistrib} and the age $\tau_{\rm{max}}$ refer
to this component only.

\vspace{0.2cm}
\item[(II)] The `low-metallicity' component
\vspace{0.1cm}

For this component we use only the isochrones with the four lowest metallicities from
Table\,\ref{tab8}:

Stars represented by the isochrones with $\hbox{[Fe/H]}=-0.95$ and 
$-0.78$ with contributions of $\sim0.5\%$ and $\sim1.0\%$ have ages $\tau \in
(\tau_{\rm{max}}, \tau_{\rm{max}}+0.6\Gyr$).

Stars represented by $\hbox{[Fe/H]}=-0.65$ with a total contribution of $\sim1.5\%$;
two thirds of these have ages $\tau \in (\tau_{\rm{max}}-0.5\Gyr$,$
\tau_{\rm{max}}+0.1\Gyr$)
and one third has ages $\tau \in (\tau_{\rm{max}}-5\Gyr$,$
\tau_{\rm{max}}-0.5\Gyr$).

Stars represented by $\hbox{[Fe/H]}=-0.47$ with a contribution of $\sim1.0\%$  have ages 
$\tau \in (\tau_{\rm{max}}-1\Gyr$,$ \tau_{\rm{max}}$). There is also a significant contribution
from stars with this metallicity to the `thin disc' component (cf. Section \ref{subsec:Z}).

It would seem desirable to assign a velocity dispersion as high as the one of the `thick disc'
to this component, however this significantly diminishes the quality of the fits. We thus
model both components with a single disc heating rate as described in the following Section and
discuss this result in Section \ref{subsec:QG}.

\end{itemize}

\subsection{The disc heating rate}\label{subsec:heating}

We model the velocity dispersion $\sigma^2$ of a group of stars with a known
distribution in age by
 \begin{equation}
 \sigma^2=\frac{\int_0^{\tau_{\rm{max}}}\d\tau\,({\d N}/{\d\tau})\sigma^2(\tau)} 
{\int_0^{\tau_{\rm max}}\d\tau\, ({\d N}/{\d\tau})}.
\end{equation}
 A simple and often used model \citep[e.g.][\S8.4]{GD2}, that we will
mainly consider here is 
 \begin{equation}
\sigma(\tau)=v_{10}\left(\frac{\tau+\tau_1}{10\Gyr+\tau_1}\right)^{\beta},
\end{equation}
 where $v_{10}$ and $\tau_1$ characterise the velocity dispersion at 10 Gyr
and at birth and $\beta$ describes the efficiency of stochastic
acceleration.
However, \citet{Quillen} argued for saturation of disc heating
after $\sim3\Gyr$ and an abrupt increase in velocity dispersion at
$\sim9\Gyr$, which they connected to the formation of the thick disc.
Therefore we also consider the following model
 \begin{equation}\label{QGsigma}
 \sigma(\tau) = \left\{
              \begin{array}{lll}
                   {\displaystyle
		   v_s\left(\frac{\tau+\tau_1}{T_1+\tau_1}\right)^{\beta}} & \rm{for} & \tau<T_1\\
                   v_s & \rm{for}& T_1<\tau<T_2\\
                   \eta v_s& \rm{for}&\tau>T_2,
              \end{array}
       \right.
\end{equation}
 where $T_1$ and $T_2$ are the times for the occurrence of saturation and the
abrupt increase, $v_s$ and $\eta v_s$ are the saturation velocity dispersions
for the thin and the thick disc and $\beta$ and $\tau_1$ are as above.

Our velocity dispersion data from Section \ref{sec:kinematics} are not for an infinite cylinder,
but for a volume-limited Hipparcos sample. As a coeval population heats, it
will spread in $z$ and the fraction of its stars that contribute to the
\textit{Hipparcos} sphere will drop. So the contribution of this population
to the measured dispersion will be less than that of a younger population. We
resolve this problem by introducing a weighting factor $F(\tau)\le1$ such
that a population's contribution to the measured dispersion is proportional
to
 \begin{equation}
\frac{\d N}{\d\tau}_{\rm heating}\equiv F(\tau)\,\frac{\d N}{\d\tau}_{\rm cylinder}.
\end{equation}
 To estimate $F(\tau)$ we consider the conservation of the number
of stars born in a certain time interval $\d\tau$ in the approximation that
we can neglect radial mixing. Then, as the population heats
and spreads in $z$, its central density drops by a factor $F(\tau)$, which is
given by
 \begin{equation}
F(\tau)\int_0^{\infty} \d z \,
\exp\left(-\frac{\Phi(z)}{\sigma_W^2(\tau)}\right)=\rm{constant}.
\end{equation}
The contribution of a population to
any sphere around the sun can now be obtained from equations
(\ref{defsNcyl}) and (\ref{givesnuzero}). For our first fit of the model to
the data we take $F(C,\tau)=1$. The resulting function $\sigma_W(\tau)$ is
used to determine $F(C,\tau)$, a new fit is made and $F$ is redetermined.
This sequence of operations is rapidly convergent.

\vspace{0.15cm}
 The Levenberg--Marquardt non-linear least-squares algorithm \citep{Press} is
used to minimise the $\chi^2$ of the fits to the data for $\sigma(B-V)$ and
$\d N/\d C(B-V)$. The parameters adjusted are (for the standard disc heating
and SFR model) $\alpha$, $\beta$, $\gamma$, $\tau_{\rm{max}}$, $\tau_1$ and
$v_{10}$.  Data at $B-V<0$ are discarded for fear that the sample of young
stars is kinematically biased. As stated above, the stars are binned in
sliding windows of 500 stars, a new one every 100 stars. It might seem
desirable to use only statistically independent bins (every fifth bin), but
then the results turn out to depend on which subset of bins is used. The
compromise used was to use every third bin, which reduces the degradation of
the information available about the colours at which dispersions change, at the
price of yielding values of $\chi^2$ that are slightly too low.

\begin{table*}
	\centering
	\caption{The parameters for the best fits to the data at different
metallicity distributions and the fit quality $\chi^2$. The metallicity
distribution of the 'thin disc' component is characterised by the effective
weights of the isochrones with the ten highest values of $\hbox{[Fe/H]}$ in Table\,\ref{tab8} }
  \label{tab1}
		\begin{tabular}{lc|cccccc|c}
\hline
$\Delta_{\rm[Fe/h]}$ & effective weights $W_i$  & $\alpha$ & $\beta$ & $\gamma$ & $\tau_{\rm{max}}$ & $\tau_1$ & $v_{10}$ & $\chi^2$ \\
&`thin disc' comp.&&&&&&&\\
 dex&low $\rightarrow$ high $Z$ in       \%   &          &         &  Gyr$^{-1}$& Gyr          & Gyr      & km\,s$^{-1}$&\\
\hline
&with low met. tail&&&&&&&\\
\hline
0&(0.4, 3.5, 10.0, 16.0, 18.5, 19.5, 16.0, 11.0, 4.4, 0.7)& $-2.559$& 0.356 & 0.117 & 12.557 & 0.187 & 55.187 & 1.05\\
0.1&(0.0, 1.5,  8.0, 17.0, 22.0, 23.0, 17.0,  9.0, 2.5, 0.0)& $-2.549$& 0.349 & 0.117 & 12.602 & 0.149 & 55.179 & 1.04\\
0.12&(0.0, 0.7,  6.0, 17.0, 24.0, 26.0, 17.5,  7.5, 1.3, 0.0)& $-2.571$& 0.350 & 0.115 & 12.601 & 0.148 & 55.232 & 1.06\\
\hline
&without low met. tail&&&&&&&\\
\hline
0&(0.4, 3.5, 10.0, 16.0, 18.5, 19.5, 16.0, 11.0, 4.4, 0.7)& $-2.519$& 0.385 & 0.119 & 12.606 & 0.261 & 57.157 & 0.92\\
0.1&(0.0, 1.5,  8.0, 17.0, 22.0, 23.0, 17.0,  9.0, 2.5, 0.0)& $-2.673$& 0.375 & 0.121 & 11.980 & 0.201 & 57.588 & 0.94\\
0.12&(0.0, 0.7,  6.0, 17.0, 24.0, 26.0, 17.5,  7.5, 1.3, 0.0)& $-2.610$& 0.376 & 0.130 & 11.782 & 0.190 & 57.975 & 1.03\\
\hline
		\end{tabular}
\end{table*}

\section{The results of the fits}\label{sec:fits}

As described in BDB00, there are correlations between the parameters which
limit the usefulness of the formal errors on the parameters. So we present
the results in the following form: we first show the influence of different
metallicity weightings on the results and after standardising on a plausible
configuration demonstrate the possible range of
each parameter by showing the results obtained for fixing it to certain
values and leaving the other parameters free.
We generally use the total velocity dispersion for our models.  The components
of $\sigma$ are studied in Section \ref{subsec:sigmaz}.

\subsection{The influence of the metallicity weighting}\label{subsec:Z}

It is interesting to study how our results depend on the weights we assign to
the sequence of the ten isochrones in Table\,\ref{tab8} with the highest
metallicities, which together represent the `thin disc'.  As explained in
Section \ref{subsec:agemet} we consider three values for the observational
scatter in [Fe/H]: 0, 0.1 and $0.12\dex$.

\begin{figure}
\includegraphics[width=.9\hsize]{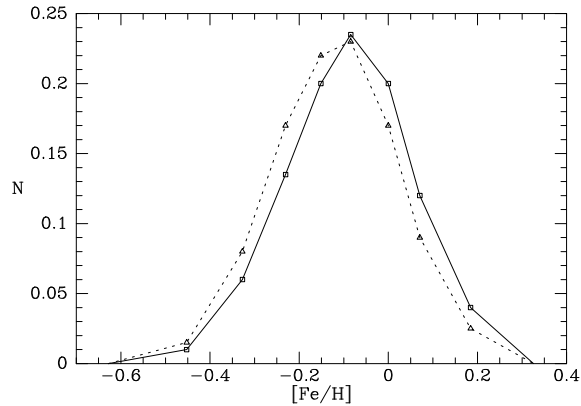}
 \caption{The full curve joins the intrinsic weights $w_i$ of the isochrones
$i=3,\ldots,12$ for the `thin disc' component that yield the effective weights $W_i$ of the isochrones in a
magnitude-limited sample like that of GCS2 that are joined by the dashed
curve. After convolution by a Gaussian
distribution of observational errors with dispersion $\Delta_{{\rm
[Fe/H]}}=0.1\dex$
the $W_i$ are consistent with the metallicity distribution of GCS2.}
\label{newfig}
\end{figure}

GCS2 give the metallicity distribution of F and G dwarfs near the sun, which
is a biased measure of the relative numbers of stars formed with each
metallicity because stars of a given mass but different metallicities have
different lifetimes and luminosities and therefore probabilities of entering a
magnitude-limited sample. Fortunately our model simulates this bias; we
simply have to choose the weights of the isochrones such that the
contribution of each chemical composition to the modelled magnitude-limited
sample agrees with the distribution of metallicities in GCS2. Let $w_i$ be
the proportion of the SFR which goes into stars of the $i$th chemical
composition. Then the effective weight of this composition is $W_i=F_iw_i/W$,
where $W\equiv\sum_i F_iw_i$ and
 \begin{equation}
F_i\equiv\int\d(B-V)\,\int\d\tau\,\left({\d
N_i\over\d\tau}\right)_{B-V},
\end{equation} 
 with $(\d N_i/\d\tau)\d\tau$  the number of stars
in the given colour and age range that we would have if the whole disc
consisted of stars of the $i$th chemical composition. The $w_i$ are chosen
such that after convolution by an appropriate Gaussian distribution of
measuring errors  the effective weights $W_i$ agree with the metallicity distribution determined by
GCS2.  The broken curve in \figref{newfig} shows the effective weights $W_i$ that
correspond to the intrinsic weights $w_i$ of the isochrones, which are
shown by the full curve -- note that neither curve looks Gaussian because the
bins have varying widths. We see that the effective weights are biased
towards low metallicities relative to the intrinsic weights $w_i$.

The adopted metallicity distribution of the `thin disc' component hardly affects 
the results of the best fit. All parameters and the fit quality are very stable,
as is shown in Table\,\ref{tab1}.
As the construction of the low-metallicity component is not straight-forward,
we also tested models without this component. The results are also presented
in Table\,\ref{tab1}. For those models the age $\tau_{\rm max}$ depends strongly
on the metallicity distribution in the sense that the stronger the
low-metallicity tail of the distribution, the larger is the  recovered value of
$\tau_{\rm max}$.  Consequently, $\tau_{\rm max}$ decreases as we
increase our estimate of the errors $\Delta_{\rm FeH}$ in the GCS
metallicities because the larger the measurement errors, the narrower the
distribution of intrinsic metallicities used in the model.  
The models without a low-$Z$ component generally show better fit qualities than the two-component models,
but broadly speaking all fits are in the same quality range.
The old, low-metallicity component is constrained to a relatively small colour 
interval, whereas for the one-component model, the low-metallicity stars cover
all ages and thus a wider colour interval, resulting in a better fit quality.
All models with a significant low-metallicity component show very similar ages
$\tau_{\rm max}\sim12.5$Gyr.

All the ages in Table~\ref{tab1} are larger than the value $11.2\pm0.75\Gyr$
obtained by BDB00, principally because the reduction in the estimate of
reddening shifts Parenago's discontinuity to the red, where the average age
of stars is higher. The age could be brought down by lowering the helium
abundance of the low-$Z$ isochrones, which would shift the isochrones to redder colours.
However, values of $Y$ significantly
below the value of $Y_{\rm p}$ from WMAP would be required to reduce
$\tau_{\rm max}$ appreciably.  Interestingly, \citet{Casagrande} found very
low Helium abundances for nearby K-dwarfs using Padova isochrones.

For the following sections we decided to use two-component models  with a
`thin-disc' metallicity distribution 
corresponding to measurement errors of $\Delta_{\rm FeH}=0.1\dex$ and an intrinsic
dispersion of $0.14\dex$.

\vspace{1cm}

\begin{table}
	\centering
	\caption{The effect of varying $\beta$ on the fits. For $\alpha$ and $\tau_{\rm{max}}$
	upper limits of $-2.400$ and 13.000 Gyr were set. }
  \label{tab3}
		\begin{tabular}{c|c|cccc|c}
\hline
 $\alpha$ & $\beta$ & $\gamma$ & $\tau_{\rm{max}}$ & $\tau_1$ & $v_{10}$ & $\chi^2$ \\
         &fixed        &  Gyr$^{-1}$& Gyr          & Gyr      & km\,s$^{-1}$&\\
\hline
$-2.400$& 0.250 & 0.141 & 13.000 & 0.001 & 48.257 & 3.78\\
$ -2.445$& 0.300 & 0.119 & 13.000 & 0.001 & 52.615 & 1.29\\
$-2.549$& 0.349 & 0.117 & 12.602 & 0.149 & 55.179 & 1.04\\
$ -2.643$& 0.420 & 0.112 & 12.281 & 0.486 & 57.449 & 1.19\\
$ -2.722$& 0.500 & 0.120 & 11.605 & 0.978 & 58.963 & 1.58\\

\hline
		\end{tabular}
		\end{table}

		\subsection{Varying $\beta$}\label{subsec:beta}

\begin{figure*}

\hspace{6.73cm}
\scalebox{0.8}{\includegraphics[angle=270]{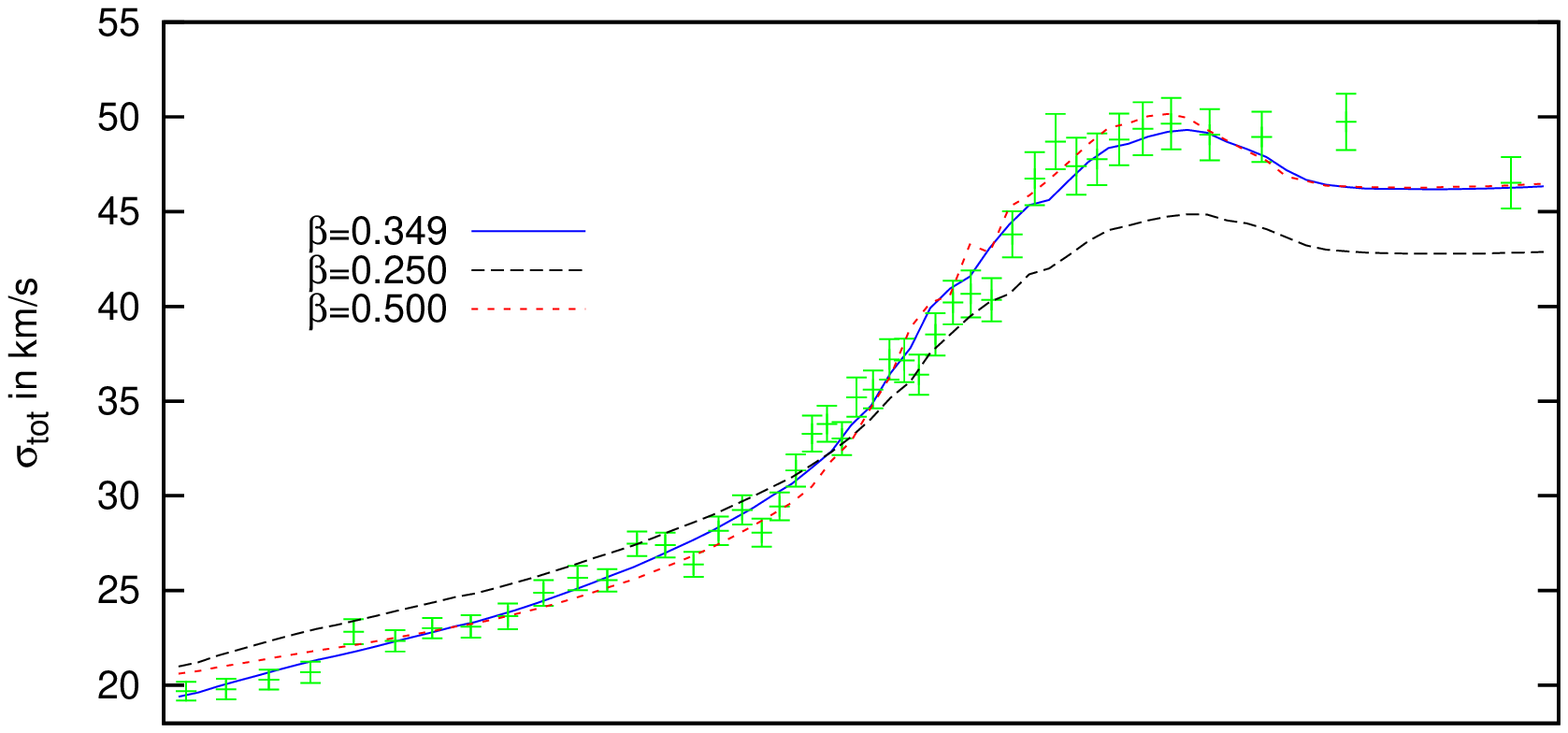} }\\
\vspace{ -0.82cm}

\hspace{-0.22cm}\scalebox{0.813}{\includegraphics[angle=270]{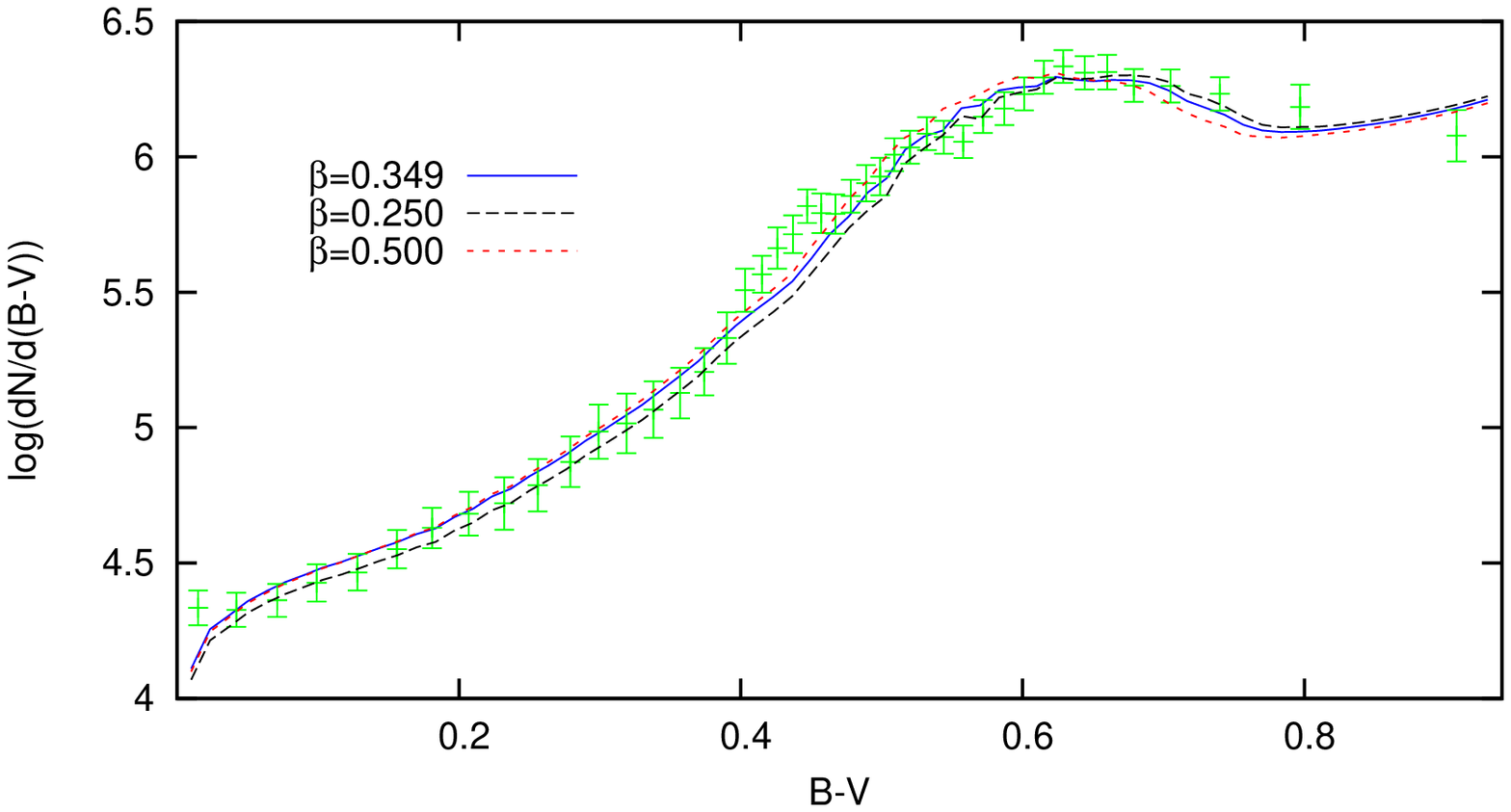} }
   \caption{Fits to the data for different values of $\beta$. \textit{Upper
panel:} Total velocity dispersion, \textit{Lower panel:} Projected number in the
solar cylinder per colour bin. \textit{Green}: Input Data, \textit{Long
dashed, black}: $\beta=0.250$, \textit{Blue}: $\beta=0.349$, \textit{Short
dashed, red}: $\beta=0.500$.}
   \label{fig11}
   \end{figure*}

Table\,\ref{tab3} shows the results of fixing $\beta$, the exponent in the
heating rate, and Fig.\,\ref{fig11} shows the fits obtained with the best
value ($\beta=0.349$) and fits for values of $\beta$ that are just too large
($\beta=0.500$) and clearly too small ($\beta=0.250$) to be acceptable.  For
$\beta=0.250$ the dependence of $\sigma$ on $B-V$ is too flat and we obtain a
poor fit to the number counts. Judging from the results, we are able to exclude
values of $\beta\la0.28$. The fit for $\beta=0.500$  is relatively bad for the
$N$-data of red stars and the corresponding $\sigma(B-V)$ is too flat for blue stars
and too steep just before and around Parenago's discontinuity.  In view also of the large $\chi^2$, 
and the high velocity dispersion at birth
($17.6\kms$), we conclude that $\beta\geq0.50$ can be excluded.

When $\beta$ is increased, the program increases $v_{10}$ and
$\tau_1$ in an effort to keep the general shape of $\sigma(\tau)$ constant.
Since the velocity dispersion of red stars strongly constrains
$v_{10}\tau_{\rm max}^\beta$, the increase in $v_{10}$ is compensated by a
decrease in $\tau_{\rm max}$. As the isochrones have limted age ranges,
 we generally set an upper limit for $\tau_{\rm max}$ at 13.0 Gyr.
$\alpha$ and $\gamma$ are relatively stable for the different values of
$\beta$. $\alpha$ is only allowed to vary around the \citet{KTG} value $-2.7$
within the interval  $(-2.4,-3.0)$.
		
		\begin{figure*} 

\scalebox{0.8}{\includegraphics[angle=270]{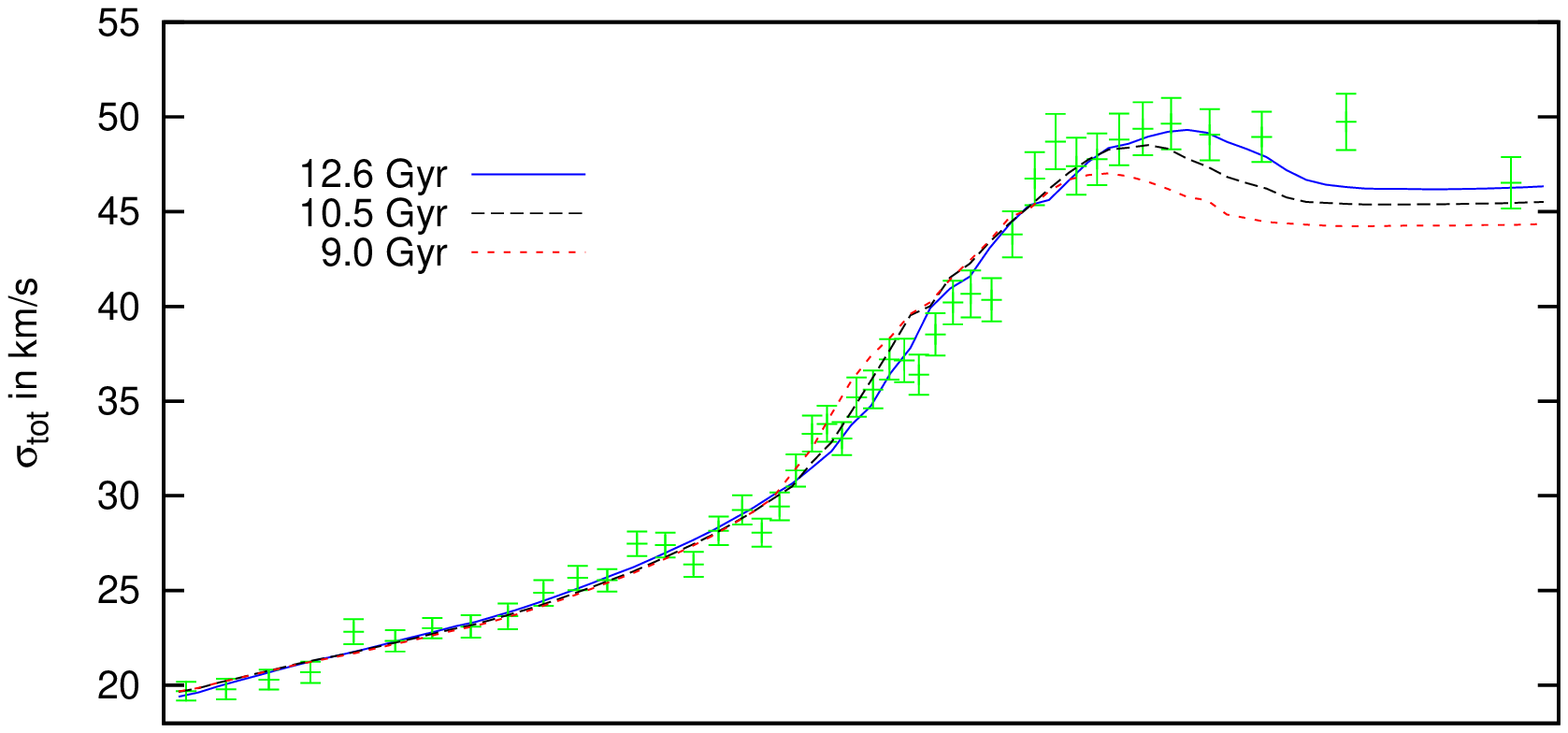} }\\
\vspace{-0.82cm}

\hspace{-0.22cm}\scalebox{0.813}{\includegraphics[angle=270]{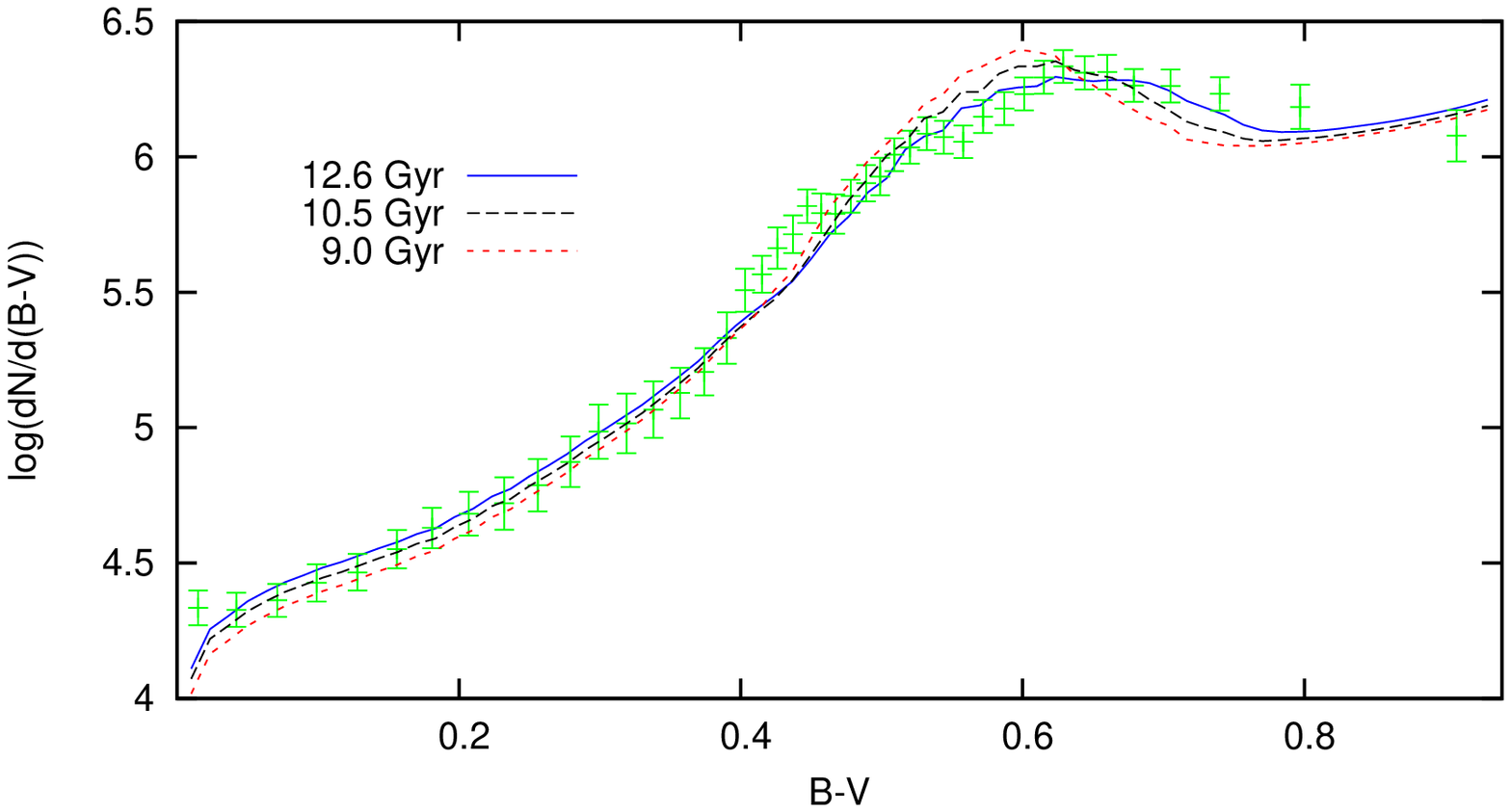} }
  
 \caption{Fits to the data for different values of $\tau_{\rm{max}}$.
\textit{Upper panel:} Total velocity dispersion, \textit{Lower panel:} Projected
number in the solar cylinder per colour bin. \textit{Green}: Input Data,
\textit{Blue}: $\tau_{\rm{max}}=12.6 $Gyr, \textit{Long dashed, black}:
$\tau_{\rm{max}}=10.5$Gyr, \textit{Short dashed, red}: $\tau_{\rm{max}}=9.0$
Gyr.}
   \label{fig12}
   \end{figure*}

	\begin{table}
	\centering
	\caption{The effect of varying $\tau_{\rm{max}}$ on the fits. For the lowest
	ages considered, $\alpha$ was fixed at $-2.400$.}
  \label{tab4}
		\begin{tabular}{ccc|c|cc|c}
\hline
 $\alpha$ & $\beta$ & $\gamma$ &$\tau_{\rm{max}}$ & $\tau_1$ & $v_{10}$ & $\chi^2$ \\
         &         &  Gyr$^{-1}$& fixed         & Gyr      & km\,s$^{-1}$&\\
\hline
$-2.510$& 0.338 & 0.112 & 13.000 & 0.103 & 54.758 & 1.07\\
$-2.549$& 0.349 & 0.117 & 12.602 & 0.149 & 55.179 & 1.04\\
$-2.503$ & 0.357 & 0.131 & 12.000 & 0.182 & 55.495 & 1.07\\
$ -2.470$& 0.363 & 0.153 & 11.500 & 0.215 & 55.499 & 1.18\\
$ -2.400$& 0.375 & 0.173 & 11.000 & 0.276 & 55.696 & 1.30\\
$ -2.400$& 0.383 & 0.195 & 10.500 & 0.307 & 55.951 & 1.43\\
$ -2.400$& 0.393 & 0.217 & 10.000 & 0.361 & 56.124 & 1.72\\
$ -2.400$& 0.404 & 0.274 & \,\,9.000 & 0.394 & 56.594 & 2.55\\
\hline
		\end{tabular}
\end{table}

\subsection{Varying the age}\label{subsec:age}

We now examine the effect of optimising the other parameters for fixed age
$\tau_{\rm{max}}$. The favoured ages are higher than in BDB00 and
surprisingly large in relation to the accepted age of the universe $\tau_{\rm
H}=13.7$\,Gyr.
So we fixed $\tau_{\rm{max}}$ at values $\ge9\Gyr$ to see if acceptable younger
models could be found. Table\,\ref{tab4} shows the results and Fig.\,\ref{fig12} 
displays three fits to the data.

$\gamma$ increases with decreasing age as the ratio of old stars to young
stars is an important constraint influencing both $\sigma(B-V)$ and $\d N/\d
C$. $\alpha$ slightly increases with increasing $\gamma$, but is relatively
stable. As explained above, low ages are associated with higher values of
$\beta$, $\tau_1$ and $v_{10}$.

Judging from the values of $\chi^2$ listed in Table~\ref{tab4}, ages above
$11.5\Gyr$ are certainly accepted and ages of $10\Gyr$ and lower can be
excluded. From Fig.\,\ref{fig12} we see that the problem with low ages lies in
the reddest bins, in both the $N$ and the $\sigma$ data. The fit for $9\Gyr$  is
definitely unable to represent the data and such low ages have to be
excluded. For $10.5\Gyr$, $\chi^2$ is still reasonable but the fit redwards
of $B-V=0.55$ looks bad. In view of the development of $\chi^2$ for ages
between 11.5 and $10\Gyr$, we can set a conservative lower age limit $\sim10.5\Gyr$.

\begin{table*}
	\centering
	\caption{The effect of varying the parameter characterising the
	velocity  dispersion at birth $\tau_1$ on the fits. }
  \label{tab5}
		\begin{tabular}{|cc|cccc|c|c|c}
\hline
 $\sigma_{\rm birth}$ & $\sigma_{\rm final}$&$\alpha$ & $\beta$ & $\gamma$ & $\tau_{\rm{max}}$ & $\tau_1$ & $v_{10}$ & $\chi^2$ \\
$\!\kms$&$\!\kms$ &   &     &  Gyr$^{-1}$& Gyr          & fixed      & km\,s$^{-1}$&\\
\hline
0.00&58.90&$-2.400$& 0.317 & 0.117 & 13.000 & 0.000& 54.200 & 1.11\\
8.14&59.46&$-2.519$& 0.329 & 0.116 & 12.673 & 0.030 & 55.056 & 1.09\\
 12.65&61.82&$-2.549$& 0.349 & 0.117 & 12.602 & 0.149 & 55.179 & 1.04\\
  16.04&61.36&$-2.635$& 0.416 & 0.119 & 12.096 &0.500 & 56.944 & 1.17\\
 17.27&62.22&$-2.728$& 0.462 & 0.119 & 12.020 & 0.800 & 57.483 & 1.39\\
\hline
		\end{tabular}
\end{table*}

\subsection{The velocity dispersion at birth and after $\tau_{\rm
max}$}\label{subsec:v10}

The correlations of the parameters $\tau_1$ and $v_{10}$, which determine the
velocity dispersion at birth and for the oldest stars, have already been
discussed. $v_{10}$ only varies within a narrow range, as it is strongly
constrained by the observed velocity dispersion redwards of Parenago's
discontinuity. As Table\,\ref{tab5} shows, the data allow a wide range of
$\tau_1$ values that correspond to velocity dispersions at birth between 0
and $\sim15\kms$. The effect of this parameter is quite small as the
power-law leads to a steep increase of velocity-dispersion within a time
small compared to $\tau_{\rm{max}}$. As the dispersion is already $20\kms$
for the bluest bin, the adjustment of the parameters to fit $\sigma(B-V)$ is
a minor issue.

\begin{table}
	\centering
	\caption{Fits to the data for the three different components of $\sigma$.} 
  \label{tab7}
		\begin{tabular}{|c|ccc|c}
\hline
 &  $\beta$ & $\tau_1$   & $v_{10}$ & $\chi^2$ \\
&           &   Gyr      & km\,s$^{-1}$&\\
\hline
$\sigma_U$& 0.307& 0.001& 41.899 & 0.83\\
$\sigma_V$& 0.430& 0.715 & 28.823 & 0.51\\
$\sigma_W$& 0.445& 0.001& 23.831 & 0.43\\
\hline
		\end{tabular}
\end{table}

		\subsection{Studying the components of $\sigma$}\label{subsec:sigmaz}

As the results in Section \ref{sec:kinematics} have shown, the three
components of $\sigma$ show significantly different behaviours: their ratios
vary with colour and Parenago's discontinuity is sharper in some components
than in others. Therefore we now
seek separate models for the evolution of each eigenvalue of $\sigma^2$.
Since all three models have to satisfy the same $N$ data, and the errors on
the eigenvalues of $\sigma^2$ are larger than those on $\sigma_{\rm total}$,
we fix the parameters that describe the star formation history at the
best-fit values determined above for $\sigma_{\rm total}$, namely
$\alpha=-2.55$, $\gamma=0.117$ and $\tau_{\rm{max}}=12.5\Gyr$, and determined
for each eigenvalue of $\sigma^2$ only values of the heating parameters,
$\beta$, $\tau_1$ and $v_{10}$. The data for $\sigma_W$ were represented by the
polynomial fit shown in \figref{fig4}.

Table\,\ref{tab7} shows the best-fit values of the parameter and $\chi^2$.
The latter are quite small for $\sigma_V$ and $\sigma_W$ because the formal
errors on these components are large.  Compared to $\beta_{\rm total}$,
$\beta_U=0.307$ is lower, so the heating exponents for the other two
components are higher than 0.349. Consequently, $\beta_V=0.430$ and
$\beta_W=0.445$ are significantly larger than $\beta_U$.

The three $\beta$ values are consistent with Figure \ref{fig10} in the sense
that ${\sigma_1}/{\sigma_2}$ and ${\sigma_1}/{\sigma_3}$ both decrease with
increasing $B-V$ and thus increasing mean age, and ${\sigma_1}/{\sigma_3}$
decreases more steeply than ${\sigma_1}/{\sigma_2}$, indicating that
$\beta_W>\beta_V>\beta_U$, which is what we find. However, it has to be 
mentioned, that because of the rather large interval of acceptable values for
$\beta$ encountered in Section \ref{subsec:beta}, we cannot exclude $\beta_W\leq
\beta_V$.

\begin{table}

	\centering
	\caption{The effect of varying $\gamma$ at fixed $\alpha=-2.7$ on the fits. As an upper limit for 
	the age $\tau_{\rm{max}}$ we apply 13.0 Gyr.}
  \label{tab2}
		\begin{tabular}{|c|ccccc|c}
\hline
 $\alpha$ & $\beta$ & $\gamma$ & $\tau_{\rm{max}}$ & $\tau_1$ & $v_{10}$ & $\chi^2$ \\
        fixed  &         &  fixed& Gyr          & Gyr      & km\,s$^{-1}$&\\
\hline
$ -2.700$& 0.347 & 0.140 & 12.044 & 0.147 & 54.898 & 1.21\\
$ -2.700$& 0.349 & 0.120 & 12.542 & 0.144 & 55.237 & 1.07\\
$ -2.700$& 0.348 & 0.100 & 13.000 & 0.136 & 55.396 & 1.06\\
$ -2.700$& 0.359 & 0.080 & 13.000 & 0.163 & 56.280 & 1.12\\
$ -2.700$& 0.371 & 0.060 & 13.000 & 0.194 & 57.195 & 1.47\\
\hline
		\end{tabular}
\end{table}

\begin{table*}
	\centering
	\caption{The results for a SFR with two timescales. The two input
parameters are displayed on the left, the output as usual on the right,
followed by the characteristic model parameters $\vartheta$ and $\rho$.
Keep in mind the best $\chi^2$ for the chosen metallicity distribution was
1.04. }
  \label{tab9}
		\begin{tabular}{cc|cccccc|cc|c}
\hline
 $A$&$\lambda$& $\alpha$ & $\beta$ & $\gamma$ & $\tau_{\rm{max}}$ & $\tau_1$ & $v_{s}$& $\vartheta$& $\rho$&$\chi^2$ \\
&Gyr$^{-1}$ & &        &  Gyr$^{-1}$& Gyr          & Gyr      & km\,s$^{-1}$ &&&\\
\hline
$3\times10^{-12}$&3.0&$-2.512$& 0.346 & 0.095 &10.022& 0.132& 54.968& 13.2& 0.21 &1.19\\
$1\times10^{-12}$&3.0&$-2.530$& 0.346 & 0.092 &10.363& 0.137& 54.979 &13.2& 0.18 &1.10\\
$3\times10^{-13}$&3.0&$-2.527$& 0.345 & 0.098 &10.741& 0.145& 54.823 &10.3& 0.15 &1.10\\
\hline
$3\times10^{-8}$&2.0&$-2.538$& 0.349 & 0.085 &10.233& 0.147& 55.145 &9.7& 0.21 & 1.18\\
$1\times10^{-8}$&2.0&$-2.501$& 0.348 & 0.093 &10.727& 0.162& 54.821 & 7.7&0.16 &1.11\\
$3\times10^{-9}$&2.0&$-2.530$& 0.349 & 0.089 &11.333& 0.170& 54.780 & 7.6&0.14 &1.10\\
\hline
$3\times10^{-4}$&1.0&$-2.527$& 0.350 & 0.073 &10.604& 0.170& 55.058 & 5.6&0.22 &1.18\\
$1\times10^{-4}$&1.0&$-2.530$& 0.345 & 0.092 &11.416& 0.159& 54.574 & 3.2&0.14 &1.12\\
\hline
		\end{tabular}
\end{table*}

\subsection{Different approaches to the SFR}\label{subsec:SFR}

$\alpha$ and $\gamma$ are strongly correlated. A high, positive value of
$\gamma$, i.e.\ a higher SFR in the past,  increases the number of
red stars relative to blue stars. This effect can be cancelled by a relatively
flat IMF creating more blue stars.
For the investigation of different star formation histories, we thus fixed 
$\alpha$ at the \citet{KTG} value of $-2.7$ and tested different values of $\gamma$.
The best fit in Section \ref{subsec:age} showed a less steep IMF, so we expect
now to find that the model with lowest $\chi^2$ has a lower value of $\gamma$.
Table \ref{tab2} confirms that this is the case.
In all the above Sections, the SFR is decreasing: for the best fit from Section
\ref{subsec:age}, it decreases by a factor 4.4 between the beginning of star formation
until now.
Combining Tables \ref{tab2} and \ref{tab4}, factors between 2.2 and 6.7 are plausible.

For values of $\gamma\le0.1$, we had to fix the age $\tau_{\rm{max}}$ to 13.0
Gyr for the reason given above. Our models generally need a large number of
old stars. Since stars $10\Gyr$ and older move very slowly in the CMD, it is
interesting to ask whether acceptable models with a reduced lower age limit
can be obtained by permitting a high SFR early on. Moreover, a short early  period
of intense star formation is envisaged in the popular proposal that  the
thick disc formed as a result of a major accretion event $\sim10\Gyr$ ago
\citep[e.g.][]{Chiappini01}.

We thus applied a SFR as described by equation (\ref{SFRexp2}). The timescale of
the first formation epoch should be small compared to $\tau_{\rm{max}}$, so
we tried $\lambda=1$, 2 and $3\Gyr^{-1}$. The results
for different parameters A are displayed in Table \ref{tab9}.

The fraction $\rho$ of solar-neighbourhood stars that belong to the `thick
disc' is determined by $A$ and $\lambda$. Table\,\ref{tab9} gives $\rho$ for
each model -- observationally \citet{Juric} found $\rho\approx0.11$ for their
definition of the thick disc, so we used this value as a guidance in
adjusting the parameters.
We further characterise the scenarios by the ratio of the two terms at the
beginning of star formation:
$\vartheta=A\,\exp[(\lambda-\gamma)\tau_{\rm{max}}]$. Since models with very
small $A$ are indistinguishable from standard pure-exponential models, we
stopped lowering $A$ when $\chi^2$ reached a value comparable to that of the
best-fitting standard model. Significantly, no two-exponential model achieved
a lower $\chi^2$ than the best standard model.

For all scenarios, the parameters $\alpha$, $\beta$, $\tau_1$ and $v_{10}$
are very stable; the disc-heating parameters do not depend significantly on
the applied SFR, so  there is no need for a further discussion. 
Compared to its values from Section
\ref{subsec:age}, $\gamma$ has decreased for the lower ages  to compensate for
the influence of the added
`thick disc' SFR term.  The most striking difference is that the favoured ages
have decreased by $\sim1.0\Gyr$. The best results were achieved for
$\lambda\ge2.0\Gyr^{-1}$.  For the SFR considered here, the lower age limit has
to be lowered to $\sim10.0\Gyr$.

\begin{figure}
\hspace{-0.3cm}
\scalebox{0.88}{\includegraphics[angle=270]{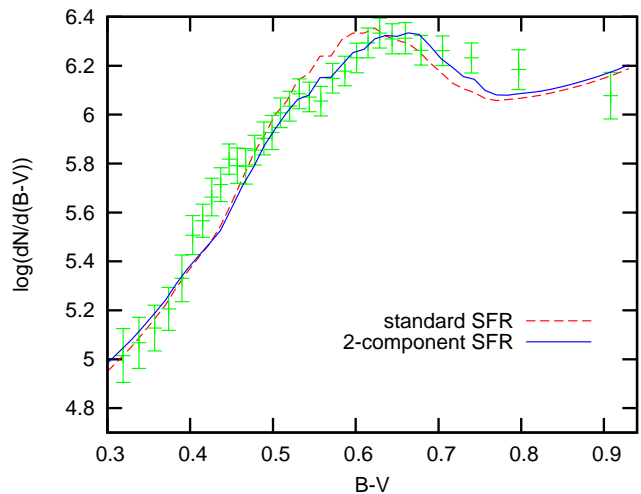} }
  
 \caption{The fits to the N-data for a standard SFR 10.5 Gyr model and a two-component ($\lambda=2.0$, $A=10^{-8}$) 10.7 Gyr model in the relevant $B-V$ interval.
}
   \label{fig16}
   \end{figure}

Fig.\,\ref{fig16} compares the fits to the $N$ data for the $10.5\Gyr$ model
of Section \ref{subsec:age} and the $10.7\Gyr$ model with $\lambda=2\Gyr^{-1}$ and
$A=10^{-8}$. It shows how the intense first period of star formation improves
the fits to the data of the red colour bins at lower disc ages.

   \Citet{RP} argue that the SFR of the solar neighbourhood is very irregular.
They also find that the SFR shows an increasing tendency, which is incompatible with our
findings. It is however interesting to ask if it is possible to achieve reasonable
fits with a smooth SFR overlaid with factor varying in time according to Fig. 8 of
\citet{RP}.
The best fit for this approach has a $\chi^2$ of 1.44 and is characterised by the following
parameters 
\begin{eqnarray} 
\alpha=-2.635        & \beta=0.354  & \gamma=0.120 \nonumber \\
\tau_{\rm{max}}=12.266 & \tau_1=0.159 & v_{10}=56.488, \nonumber
\end{eqnarray} 
which are all in the range of acceptable values as determined above. The higher
value of $\chi^2$ results from additional features, which are produced by the varying
SFR and which are incompatible with the data. 
We are, however, not able to exclude a SFR which is not smooth. As Figures \ref{fig11}
and \ref{fig12} show, there are features in the data that our models are not able to reproduce
and which might be produced by epochs of enhanced star formation.

We finally test the SFR described by equation (\ref{SFRJJ}), which was proposed
by \citet{JJ}. This SFR increases early on and then decreases. 
The best-fit model achieves $\chi^2=1.15$ with
 \begin{eqnarray}\tau_2\approx14\Gyr&& \nonumber\\
\alpha=-2.400 & \beta=0.334 & b= 7.230\nonumber\\
\tau_{\rm{max}}=12.686 & \tau_1=0.094 & v_{10}=54.392. \nonumber
\end{eqnarray}
 In this model the SFR peaked $\sim10.3\Gyr$ ago and has since decreased by a
factor $\sim4$. The fit provided by this is of a similar quality as the best
fits with the standard SFR, and the values of all comparable parameters are
little changed from when we used the standard SFR. 

\begin{center}
\begin{figure}
\hspace{-0.3cm}
\scalebox{0.9}{\includegraphics[angle=270]{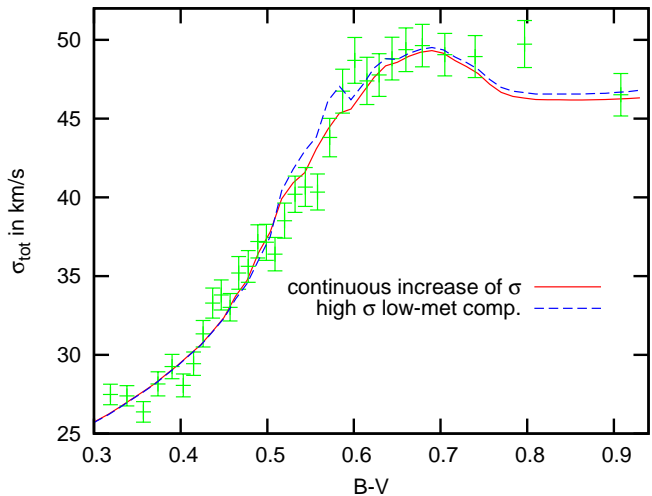} }
  
 \caption{The effect of assigning a high velocity dispersion $\sigma_{tot}=85\kms$ to the
 low-metallicity component.The red line shows the best fit from above.}
   \label{fig77}
   \end{figure}
\end{center}

\begin{table*}
	\centering
	\caption{The results for the assumption of a saturation of disc
heating. The three input parameters are displayed on the left, the output as
usual on the right.  Keep in mind the best $\chi^2$ for the chosen metallicity
distribution was 1.04. An upper age limit was set at 13.0 Gyr.}
  \label{tab6}
		\begin{tabular}{ccc|cccccc|c}
\hline
 $T_1$&$T_2$&$\eta$& $\alpha$ & $\beta$ & $\gamma$ & $\tau_{\rm{max}}$ & $\tau_1$ & $v_{s}$ & $\chi^2$ \\
Gyr&Gyr & & &        &  Gyr$^{-1}$& Gyr          & Gyr      & km\,s$^{-1}$&\\
\hline
5.0&--&1.0&$-2.493$& 0.588 & 0.114 &13.000 & 0.845& 49.468 & 1.69\\
6.0&--&1.0&$-2.413$& 0.477 & 0.117 &13.000 & 0.558& 50.245 & 1.37\\
7.0&--&1.0&$-2.480$& 0.429 & 0.119 &12.556 & 0.396& 52.044 & 1.17\\
\hline
3.0&9.0&1.50&$-2.564$& 1.986 & 0.126 &12.539 & 5.801& 41.004 & 1.51\\
3.0&9.0&1.75&$-2.509$& 0.484 & 0.132 &12.003 & 0.488& 38.483 & 1.45\\
3.0&9.0&2.00&$-2.509$& 0.458 & 0.151 &11.708 & 0.489& 37.213 & 1.67\\
\hline
3.0&8.0&1.50&$-2.448$& 0.606 & 0.127 &12.017 & 0.915& 38.929 &1.37\\
4.0&9.0&1.60&$-2.457$& 0.367 & 0.131 &12.043 & 0.205& 40.606 & 1.22\\
5.0&9.0&1.50&$-2.495$& 0.334 & 0.132 &12.046 & 0.127& 42.646 & 1.12\\
\hline
		\end{tabular}
\end{table*}

\subsection{Different Approaches to the heating rate}\label{subsec:QG}

In our models we have so far assumed that the velocity dispersion increases
continuously with age. In the context of the Galactic thick disc, an obvious
question to ask is what would be the effect of assigning a high velocity
dispersion, $\sigma_{\rm tot}=85\kms$ \citep{Bensby}, to the low-metallicity
component. \figref{fig77} shows that this procedure produces an additional
feature in the curve $\sigma_{\rm tot}(B-V)$ that underlines again the narrow
colour interval to which the low-metallicity component is confined.  The
reason for this is that the thick disk is not only metal-poor and old, but
has a more complex structure \citep{Ralphb}.

In this context it is also interesting to study whether a saturation of disc
heating,
or a merger event producing a discontinuity in the time evolution of
$\sigma$, or a
combination of the two is compatible with the data. We adopt the alternative model (\ref{QGsigma})
of $\sigma(\tau)$. This model introduces three  additional parameters, $T_1$
when heating saturates,
$T_2$ when there is a step increase in $\sigma$ and $\eta$, the scale of that
increase. Table\,\ref{tab6} provides an overview of results obtained by
fixing these parameters.

We start by looking at a pure saturation of disc heating ($\eta=1$) with
$T_1=5$, $6$ and $7\Gyr$. Intuitively, this model is incompatible with the
phenomenon of Parenago's discontinuity unless $T_1\ga7\Gyr$, and that is
what the results confirm. For $\eta=1$ we can firmly exclude $T_1\le5\Gyr$.

\begin{figure}
\hspace{-0.2cm}
\scalebox{0.88}{\includegraphics[angle=270]{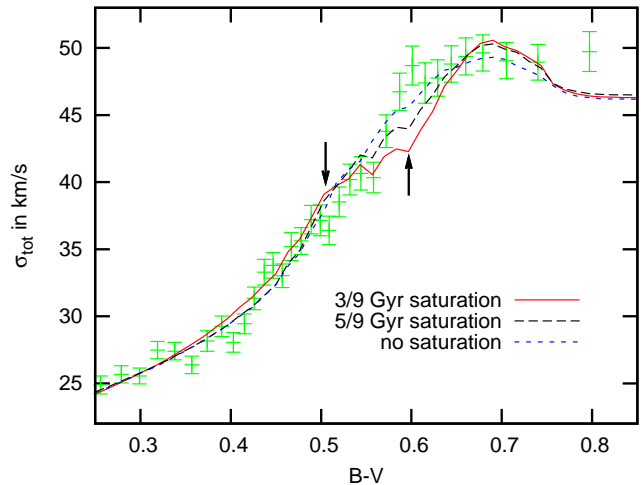} }
  
 \caption{The fits for two different saturation scenarios with ($T_1$, $T_2$,
$\eta$) = ($3\Gyr$, $9\Gyr$, 1.5) (red) and ($5\Gyr$, $9\Gyr$, 1.5) (black, long
dashed). The blue, short dashed curve is the best fit without saturation. The
black arrows indicate the two additional discontinuities in the $3/9\Gyr$
scenario.  }
   \label{fig14}
   \end{figure}

Consider now the proposal of \citet{Quillen} that $T_1=3\Gyr$, $T_2=9\Gyr$
and $\eta\ga1.5$.  Our best fit, obtained with $\eta=1.75$, is shown by the
red curve in Fig.\,\ref{fig14}. The curve of $\sigma(B-V)$ moves from the
high to the low side of the data around $B-V=0.55$ as a result of two changes in
slope at the points marked by arrows; at these points the main-sequence age
corresponds to $T_1$ and $T_2$. This figure and the high $\chi^2$ given in
Table~\ref{tab6} exclude this model.

\begin{figure}
\hspace{-0.2cm}
\scalebox{0.88}{\includegraphics[angle=270]{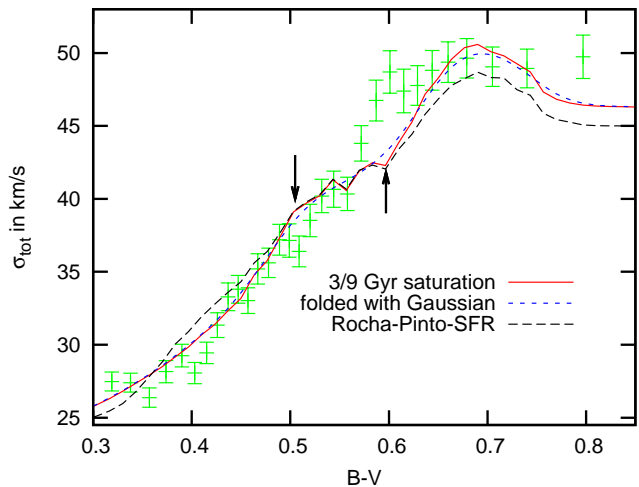} }
  
 \caption{Attempts to smooth out the discontinuities in the ($3\Gyr$,
$9\Gyr$, 1.75) $\sigma$-plot (red curve). A) Smooth SFR overlaid with a
variation in time according to \citet{RP} (black, long dashed). B) Folded
with a Gaussian with a dispersion equivalent to the error in colour $B-V$
(blue, short dashed) }
   \label{fig22}
   \end{figure}

To see whether the unwanted features associated with $T_1$ and $T_2$ could be
washed out by an irregular star-formation history rather than the smooth one
used in our models, we modelled the data using $T_1=3\Gyr$, $T_2=9\Gyr$,
$\eta=1.75$ and overlaying our smooth SFR with a factor varying with time
according to Fig.~8 of \citet{RP}.  The resulting fit
to the data, shown by the long-dashed line in \figref{fig22}, is even worse
than that shown in \figref{fig14}. Another possibility is that large errors in
$B-V$ ($\sim0.023\,$mag) could smooth away the unwanted discontinuities. To
test this hypothesis we folded $\sigma(B-V)$ for the ($3\Gyr$, $9\Gyr$, 1.75)
model with a Gaussian of $0.023\mag$ dispersion.  Still the discontinuities
did not disappear (short-dashed curve in \figref{fig22}).

Acceptable models can be found by allowing $T_1$ to approach $T_2$ with
corresponding adjustment to $\eta$. For example, the fit provided by the
($5\Gyr$, $9\Gyr$, 1.5) model is shown as the black curve in
Fig.\,\ref{fig14}: the fit to $\sigma(B-V)$ is clearly better than that for
($3\Gyr$, $9\Gyr$, 1.75) model and nearly as good as the best fit
without saturation (blue curve). Overall one has the
impression that models with $T_1$ close to $T_2$ are merely approximating a
power-law dependence with an appropriate step.

\subsection{Why does $\sigma$ decrease redwards of the
discontinuity?}\label{subsec:reddrop}

\begin{center}
\begin{figure}
\hspace{-0.3cm}
\scalebox{0.9}{\includegraphics[angle=270]{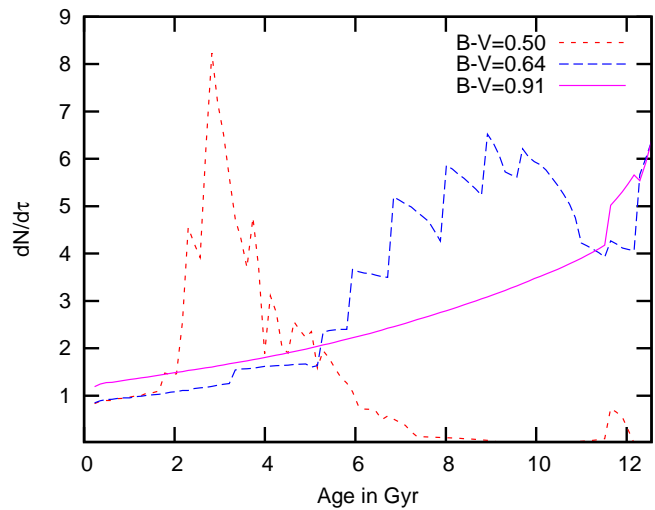} }
  
 \caption{The age distribution of the $12.6\Gyr$ fit from Section \ref{subsec:age}
for three mean colours.  }
   \label{fig15}
   \end{figure}
\end{center}

All plots of $\sigma(B-V)$, whether for models or data, show a
counterintuitive decrease in sigma for the reddest bins: the velocity
dispersion in our models increases with age, so decreasing $\sigma$ implies a
decrease in mean age as one moves redwards of the discontinuity.
\figref{fig15} shows model age distributions at $B-V\approx0.50$, $0.65$ and
$0.90$. For the reddest colour we see only a smooth increase in numbers with
age, reflecting the declining SFR, and a feature at high ages, resulting from the 
low-metallicty component.  At the bluer colours we have strong
features.  From equation (\ref{xx}) we see that they must reflect wider mass
intervals yielding stars of the given colour at a certain time. The explanation
for this is that in the vicinity of the turnoff, the isochrones run almost
vertically and the colours of stars become nearly independent of mass for a
significant range of masses. In \figref{fig15} the blue age distribution for
$B-V\approx0.65$ shows eight peaks produced in this way, one from each of the
eight metallicities used for the `thin disk' component and also a low-metallicity
feature at high ages. Because the dominant peaks all lie at ages higher than
$5\Gyr$, they raise the average age of stars in this colour bin above that of
the bin for $B-V=0.91$, which contains only stars too low in mass to have
reached the turnoff.

\section{Conclusion}\label{sec:conclude}

We have updated the work of DB98 and BDB00 to take advantage of the reworking
of the \textit{Hipparcos} catalogue by \cite{van Leeuwen}, the availability
of line-of-sight velocities from the Geneva--Copenhagen survey, and
significant improvements to the modelling. The latter include updated
isochrones, better treatments of interstellar reddening, the selection
function and the age-metallicity relation, and exploration of a wider range of histories of star-formation and
stellar acceleration.

We have redetermined the solar motion with respect to the
LSR. The new value (eq.~1) is very similar to that of DB98 but has smaller error
bars. We have redetermined the structure of the velocity ellipsoid as a
function of colour. Again the results differ from those of DB98 mainly in the
reduced error bars.

The most striking change in results compared to BDB00 is an increase in the
minimum age of the solar neighbourhood $\tau_{\rm max}$. This increase is
largely due to an improved treatment of interstellar reddening, which was
overestimated by BDB00.  Models in which the SFR is $\propto\exp(\gamma\tau)$
fit the data best at $\tau_{\rm max}\approx12.5\Gyr$ and favour $\tau_{\rm
max}>11.5\Gyr$; these models yield a lower age limit of $10.5\Gyr$. Models in
which the SFR is a double exponential corresponding to the formation of the
thick and thin discs favour ages in the range $10.5-12\Gyr$ and yield a lower
limit of $10.0\Gyr$. For either model of the SFR, the lower age limit of
BDB00, $9\Gyr$, can be excluded. Our estimates are in good agreement with the
$12\Gyr$ age derived by \cite{JJ} and the $11.7\Gyr$ obtained by
SB09a.  It is also in agreement with the individual stellar ages of
GCS and GCS2, which include a significant fraction of ages between 10 and 15
Gyr. The study of Galactic evolution in a cosmological context by
\citet{Hernandez} yields $\tau_{\rm max}=11\Gyr$, while that of \citet{Naab}
yields $\tau_{\rm max}\sim10\Gyr$, our lower limit.  Thus there is
a considerable body of evidence that the solar neighbourhood started forming
a remarkably long time ago.

The strong correlation between the parameters $\alpha$ and $\gamma$ that
characterise the IMF and SFR, which was discussed by \citet{Haywood} and
BDB00, limit what we can say about the IMF and SFR.  We use a \citet{KTG}
IMF and find that the SFR is decreasing: the factor by
which it decreases from the beginning of star formation until now is found to lie
between $\sim6.5$ and $\sim2.5$.  This conclusion agrees with the findings of
\citet{Chiappini} and of \cite{JJ}, who used a non-exponential time
dependence of the SFR, and what we find when the SFR is modelled by a sum of
exponentials in time.

Our conclusion regarding the time dependence of the SFR conflicts with the
finding of BDB00 that the SFR was essentially flat for a Salpeter IMF because
the introduction of variable scale heights decreases the visibility of red
stars relative to blue stars, so the models have to predict the existence of
a higher fraction of red stars than formerly. It also conflicts with the
conclusion of SB09a that the SFR is only mildly decreasing for a
Salpeter IMF because in their models many of the old stars in the solar
neighbourhood are immigrants from smaller radii while immigration of young
stars is negligible. \citet{Hernandez} conclude that the SFR increases in the
first $\sim4\Gyr$ and then decreases by a factor 2 until now, while
\citet{RP} argue that the Milky Way disc has a generally increasing but very
irregular SFR. Neither picture is compatible with our results. There are
similar conflicts with the conclusions of \citet{BN} that the SFR is broadly
increasing and of \citet{Naab} that the SFR increases early on and has been
roughly constant over the last $\sim5\Gyr$.

We can exclude the scenario of \citet{Quillen} that disc heating saturates
after $3\Gyr$ but at $9\Gyr$ the velocity dispersion abruptly increases by a
factor of almost 2. However, we are not able to exclude a later saturation of
disc heating (after $\ga4\Gyr$) that is combined with an abrupt increase in
dispersion more recently than $9\Gyr$ ago.  Similarly, \citet{Seabroke} find that a
saturation at $\ga4.5\Gyr$ is not excluded and that there is ``extremely
tentative'' evidence of an abrupt feature in the age-velocity dispersion at
$8\Gyr$. Nevertheless, nothing in the data calls for early saturation of disc
heating, and scenarios that include it yield worse but formally acceptable
fits to the data.

Our favoured value $\beta=0.35$ for the exponent that governs the growth of
$\sigma_{\rm total}$ is in perfect agreement with the findings of GCS
(0.34) and BDB00 (0.33). Moreover, the value of GCS3 (0.40) is also still in
the range allowed by our models. The classical value of $\frac{1}{2}$
\citep{Wielen} yields rather bad fits and has to be regarded as the very
upper limit for $\beta$. 

For $\sigma_U$ we find $\beta\approx0.31$, the same value as in GCS
(0.31), but lower than that of GCS3, whose 0.39 is yet not out of range. For
$\sigma_V$ we find $\beta\approx0.43$, which is higher than the GCS and GCS3
values of 0.34 and 0.40. However, had they not ignored their oldest bins of
stars, they would have obtained a larger value for $\beta$. For $\sigma_W$ we
find $\beta\approx0.45$, in good agreement with GCS (0.47) and lower
than the value of GCS3 (0.53), but higher than the 0.375 favoured by
\citet{JJ}.

The values and time dependencies of the ratios $\sigma_1/\sigma_2$ and
$\sigma_1/\sigma_3$ plotted in \figref{fig10} provide important clues to the
still controversial mechanism of stellar acceleration. The original proposal
\citep{Spitzer} was that acceleration is a result of stars scattering off
gas clouds. This process leads to characteristic axial ratios of the velocity
ellipsoid.  \citet{Sellwood} has recently redetermined these ratios and finds
$\sigma_1/\sigma_2=1.41$ and $\sigma_1/\sigma_3=1.61$, both of which are
smaller than \figref{fig10} implies. This result suggests that scattering by
spiral arms, which increases $\sigma_1$ and $\sigma_2$ but not $\sigma_3$,
plays an important role \citep{Jenkins}. Further work is needed to
quantify this statement in the light of Sellwood's recent work.

Another indication that scattering by clouds is not alone responsible for
acceleration is the finding of \citet{Hanninen} that in simulations of disc
heating by clouds, $\beta=(0.21\pm0.02)$, a value that we have excluded. For
a disc heated by scattering off a combination of gas clouds and massive black
holes in the dark halo \citet{Hanninen} find $\beta=0.42$ and ${\sigma_1}/{\sigma_3}=1.59$, which
is excluded by \figref{fig10}. 

It is widely  believed that acceleration by spiral arms is responsible for
the concentration of solar-neighbourhood stars in the $(U,V)$ plane
\citep{Rab98,Dehnen00,Wu}. \cite{Seabroke} have argued that this
concentration undermines the concept of the velocity ellipsoid. However, it
remains the case that the velocity dispersions $\sigma_U$  and $\sigma_V$
increase systematically with age and it is not evident that any inconsistency
arises from modelling this phenomenon as we do here just because spiral
structure accelerates stars in groups rather than individually.

SB09a have recently argued that radial mixing plays an important role
in disc heating and causes a higher value of $\beta$ to be measured in the
solar neighbourhood than characterises the underlying acceleration process.
The origin of this effect is that stars migrate into the solar neighbourhood
from small radii where the velocity dispersions are relatively high. Since
the fraction of immigrants among the local population increases with stellar
age, this effect enhances the velocity dispersion of old stars more than that
of young stars, leading to a larger effective value of $\beta$.  Our value
for $\beta_W$ is in good agreement with the prediction of SB09a for a
\textit{Hipparcos} sample.

The models of SB09a are more elaborate than ours not only in that
they include radial migration but also in that they include chemical
evolution. However, their parameters are determined by fitting to the GCS
sample rather than the \textit{Hipparcos} sample employed here. The current
sample covers a wider range of colours and is better defined than the GCS
sample.  Moreover, SB09a made no attempt to match the kinematics of
the GCS sample but assumed the validity of the description of the
acceleration process given by BDB00; they confined themselves to the
metallicity distribution and Hess diagram of the GCS sample. Ideally one
would fit simultaneously the local kinematics, metallicity and Hess diagram.
However, given the importance of radial migration implied by the studies of
\citet{Haywood08}, SB09a and \citet{Ralphb}, such a study would ideally include a
more realistic treatment of the integrals of motion than the simple
separability of the radial and vertical motions assumed here and by
SB09a.

The cosmic SF rate peaked at redshifts $1-2$ \citep[e.g.][]{Lyetal07}, which
in the concordance cosmology corresponds to look-back times between $7.8$ and
$10.5\Gyr$. These times are $1-2\Gyr$ later than the times $\tau_{\rm max}$
at which star formation starts in our models, which is also the time at which
the local star-formation rate peaked. It is also slightly later than the mean
formation times, $\sim1.5\pm1\Gyr$, of bulges in a recent series of large
simulations of galaxy formation \citep{Scannapieco09}. On the other hand
these authors found that the mean formation times of discs at $R\sim10\kpc$
were $4\pm2\Gyr$. The analogous time for the one of our models of the solar
neighbourhood is
 \begin{eqnarray}
\overline{\tau}&=&\tau_{\rm H}
-{\int_0^{\tau_{\rm max}}\d t\,t\e^{\gamma t}\over
\int_0^{\tau_{\rm max}}\d t\,\e^{\gamma t}}\nonumber\\
&=&\tau_{\rm H}-\tau_{\rm max}+\gamma^{-1}\left(1-{\gamma\tau_{\rm
max}\over\e^{\gamma\tau_{\rm max}}-1}\right).
\end{eqnarray}
 The models listed in Table~\ref{tab1} yield values of $\overline{\tau}$ that
range between $5.9\Gyr$ and $6.4\Gyr$. These values fall at the upper end of
the times obtained by \cite{Scannapieco09}.  This overshoot may be connected
to the fact that their models have discs that are underweight by almost an
order of magnitude; in the models disc formation may be artificially
truncated. Thus our results are broadly in agreement with the results of
ab-initio simulations of galaxy formation in the concordance cosmology, even
though the age of the oldest solar-neighbourhood stars is remarkably large.
At least some of these stars will be immigrants from small radii, and if they
all are, star-formation will have started later than is implied by our values
of $\tau_{\rm max}$. It is currently hard to place a limit on the fraction of
the oldest stars that are immigrants because the models of SB09a
assume that the disc's scale length does not increase over time, as is likely
to be the case.

Thanks to adaptive optics, gas-rich discs, in which stars are
forming exceedingly rapidly, can now be studied observationally at $z\simeq2$
\citep[e.g.][]{Genzel08}, soon after the oldest solar-neighbourhood stars
formed. The discs observed are clumpy and highly turbulent, so it seems
more likely that they will turn into bulges than a  system as dynamically cold as the
solar neighbourhood. Nonetheless, studies of these systems bring us
tantalisingly close to the goal of tying together studies of `galactic
archaeology' such as ours with observations of galaxies forming in the remote
past.

\section*{Acknowledgements}
We thank Gianpaolo Bertelli for providing us with the latest isochrones and valuable comments on their
influence on our models and Ralph Sch\"onrich for valuable discussions.
 MA thanks Merton College Oxford for its hospitality during the academic year
2007/8.

\label{lastpage}
\end{document}